\documentclass[review,onefignum,onetabnum]{siamart190516}

\usepackage{lipsum}
\usepackage{amsfonts}
\usepackage{graphicx}
\usepackage{epstopdf}
\usepackage{algorithmic}
\usepackage{xcolor}
\ifpdf
\DeclareGraphicsExtensions{.eps,.pdf,.png,.jpg}
\else
\DeclareGraphicsExtensions{.eps}
\fi

\newcommand{\be}{\begin{equation} \begin{aligned} }
\newcommand{\ee}{\end{aligned} \end{equation}}

\newcommand{\Q}{\mathbf{Q}}

\newcommand{\w}{\mathbf{w}}
\newcommand{\q}{\mathbf{q}}
\newcommand{\F}{\mathbf{F}}
\newcommand{\f}{\mathbf{f}}
\newcommand{\g}{\mathbf{g}}
\newcommand{\h}{\mathbf{h}}

\newcommand{\B}{\mathbf{B}}
\newcommand{\SB}{\mathbf{S_b}}
\newcommand{\x}{\mathbf{x}}

\newcommand{\lapse}{\alpha}
\newcommand{\shift}{\beta}
\newcommand{\Lor}{W}

\newcommand{\halb}{\frac{1}{2}}

\renewcommand{\epsilon}{\varepsilon}

\newsiamremark{remark}{Remark}
\newsiamremark{hypothesis}{Hypothesis}
\crefname{hypothesis}{Hypothesis}{Hypotheses}
\newsiamthm{claim}{Claim}

\newcommand{\blue}[1]{{\leavevmode\color{black} #1}}

\usepackage{rotating}
\usepackage{pdflscape}

\newcommand{\mcol}[1]{\multicolumn{2}{|c|}{#1}}
\newcommand{\mcolN}[2]{\multicolumn{#1}{|c|}{#2}}

\newcommand{\OLdue}[0]{$\!\!\!\!\!\!\!\!\!\!\!\mathcal{O}(L_2)\!\!\!\!\!\!$}

\headers{Well balancing for general relativity}{E. Gaburro, M.J. Castro, and M. Dumbser}

\title{A well balanced finite volume scheme for \\ general relativity}

\author{Elena Gaburro\thanks{INRIA, Univ. Bordeaux, CNRS, Bordeaux INP, IMB, UMR 5251, 200 Avenue de la Vieille Tour, 33405 Talence cedex, France (\email{elena.gaburro@inria.fr}).}
\and Manuel J. Castro\thanks{Department of Mathematical Analysis, Statistics and Applied Mathematics, University of M\'alaga, Campus de Teatinos, 29071 M\'alaga, Spain (\email{mjcastro@uma.es}).}
\and Michael Dumbser\thanks{Department of Civil, Environmental and Mechanical Engineering, University of Trento, Via Mesiano 77, 38123 Trento, Italy	(\email{michael.dumbser@unitn.it}).}}

\usepackage{amsopn}

\begin{document}

\maketitle

\begin{abstract}
In this work we present a novel second order accurate {well balanced} (WB) finite volume (FV) scheme for the solution of 
the general relativistic magnetohydrodynamics (GRMHD) equations and 
the first order CCZ4 formulation (FO-CCZ4) of the Einstein field equations of general relativity, as well as the fully coupled FO-CCZ4 + GRMHD system.
These systems of \textit{first order hyperbolic} PDEs allow to study the dynamics of the \textit{matter} and the dynamics of the \textit{space-time} according to the theory of general \textit{relativity}. 

The new well balanced finite volume scheme presented here exploits the knowledge of an equilibrium solution of interest 
when integrating the conservative fluxes, the nonconservative products and the algebraic source terms, and also when performing the piecewise linear data reconstruction.
This results in a rather simple modification of the underlying second order FV scheme, which, however, being able to cancel numerical errors  committed with respect to the equilibrium component of the numerical solution,
substantially improves the accuracy and long-time stability of the numerical scheme when simulating small perturbations of stationary equilibria.  
In particular, the need for well balanced techniques appears to be more and more crucial as the applications increase their complexity. 
We close the paper with a series of numerical tests of increasing difficulty, where we study the evolution of small perturbations of accretion problems and stable TOV neutron stars. Our results show  
that the well balancing significantly improves the long-time stability of the finite volume scheme compared to a standard one. 


%
%
\end{abstract}



\section{Introduction}

The main goal of this work is to improve the long-term stability and accuracy of finite volume schemes in the context of numerical general relativity. 
The models that we consider here are the general relativistic Euler and magnetohydrodynamics (GRMHD) equations \cite{Banyuls97,Aloy1999c,Anton06,DelZanna2002,DelZanna2007,ADERGRMHD} and 
the so-called first order CCZ4 formulation (FO-CCZ4) of the Einstein field equations~\cite{Alic:2011a,Dumbser2018conformal,Dumbser2020GLM}. 
The governing PDE systems in general relativity are extremely challenging since they contain not only very complex equations with many unknowns, but some systems are also subject to involution constraints that affect hyperbolicity and thus well-posedness of the initial value problem. 
Moreover, the applications of interest in this field, such as the evolution of accretion disks, neutron stars and black holes,  
are characterized by a disparity of the involved space and time scales: 
indeed, we need to observe the \textit{long time evolution} of phenomena developing over \textit{astronomical distances}, such as gravitational waves, but whose behavior is strongly affected by physical perturbations developed at smaller scales like a binary neutron star system. 
Furthermore, it is particularly challenging to simulate small perturbations of stationary equilibria when the perturbations are of the order of the numerical discretization error. 
Hence, the main objective of this paper consists in improving long-time stability and accuracy of numerical methods for the simulation of astrophysical phenomena occurring close to stationary equilibrium solutions of the GRMHD and FO-CCZ4 equations. 
This will be achieved thanks to the introduction of modern well balancing (WB) techniques (see for example \cite{Bermudez1994,leveque1998balancing,Castro2008,Castro2017Book, Gaburro2018MNRAS,castro2020well} and the references \blue{therein}) inside the finite volume scheme that allow to preserve given equilibria of the governing PDE system \textit{exactly} at the discrete level, thus avoiding 
that small physical perturbations are masked by numerical discretization errors. 
To the best of our knowledge, this is the very first time that an exactly  well balanced finite volume scheme is proposed for the GRMHD and FO-CCZ4 equations of numerical general relativity, for a given set of relevant stationary solutions. 

\smallskip

A brief literature review on the state of the art of numerical general relativity as well as on existing well balanced techniques for computational fluid dynamics in the context of free surface shallow water flows and the Euler equations of gasdynamics with Newtonian gravity is given below. 

The GRMHD model combines the fluid description of matter with a simplified theory for electromagnetic fields in the absence of free charge carriers. 
It was studied for the first time in~\cite{wilson75,Anile1983,Russo1987} and \cite{Anile_book}, without and with electromagnetic fields. 
Substantial progress in the field has been made by the group of Iba\~nez and co-workers, who introduced the now universally adopted \textit{Valencia formulation} of general relativistic hydrodynamics and magnetohydrodynamics, see e.g. \cite{Banyuls97,Anton06,Font08,Marti2015} and references therein. 
There, the GRHD and GRMHD systems have been cast into conservation law form, which allowed to apply classical Godunov-type finite volume schemes for hyperbolic equations. Subsequently, many different solvers have been developed, 
see for example 
\cite{Baiotti04,Duez05MHD0,Anninos05c,DelZanna2007,Giacomazzo:2007ti,Anderson2008,Kiuchi2009,Bucciantini2011,Radice2012a,Dionysopoulou:2012pp,Radice2013b,Athena2016,Porth2017}. Other studies incorporate radiation transfer, like the
one proposed by~\cite{Takahashi2017}, or include the full Maxwell theory
in a resistive relativistic MHD formulation~\cite{Palenzuela:2008sf,DumbserZanotti,Dionysopoulou:2012pp,Bucciantini2013,Bugli2014,AloyCordero}.
Here we follow the approach of~\cite{ADERGRMHD}, where the gravitational field in fixed background space--time (Cowling approximation) are treated
via the use of \textit{nonconservative products} instead of the usually employed algebraic source terms, 
since gravity terms can be expressed as functions of the spatial derivatives of the lapse, 
the shift vector and the spatial metric tensor, i.e. some of the conserved variables.

Besides the evolution of the matter, the numerical integration of the Einstein field equations for the time evolution of the metric is another challenging problem. The covariant nature of the equations allows the choice of arbitrary moving curvilinear coordinates (gauge freedom), and hence first a suitable set of space-time coordinates must be chosen. 
The first step in this direction was achieved with the 3+1 (space + time) ADM formulation of~\cite{Arnowitt62unfindable}, see also~\cite{arnowitt2008republication,Baumgarte2010},  
which allowed to rewrite the Einstein field equations as an initial boundary value problem, but the proposed PDE system was not hyperbolic.   
\blue{A mixed elliptic-hyperbolic formalism, know as fully constrained formulation (FCF) was then proposed in \cite{bonazzola2004constrained} and its mathematical structure, well-posedness and uniqueness were further analyzed in \cite{cordero2008mathematical,cordero2009improved}.}
Then, among the first hyperbolic formulations was the BSSNOK  (Baumgarte-Shapiro-Shibata-Nakamura-Oohara-Kojima) formulation
\cite{Shibata95,Baumgarte99,Nakamura87}.
Subsequently, different alternative formulations were proposed, like the Z4 formulation \cite{Bona:2003fj,Bona:2003qn,Alic:2009},  
the Z4c system \cite{Bernuzzi:2009ex} and the CCZ4 in~\cite{Alic:2011a,Alic2013}. For an exhaustive overview of different models used in numerical general relativity, see \cite{Alcubierre:2008}.  Most of the previous formulations led to first order systems in time, but with first and second order derivatives in space, and were therefore not directly accessible for standard Godunov-type methods for first order hyperbolic systems. Instead, the systems were typically integrated with high order central finite difference schemes. The problem of hyperbolicity of different second order in space formulations of general relativity was discussed in \cite{Gundlach:2005ta}. The PDE system used in this paper is based on the first order CCZ4 formulation (FO-CCZ4) that has been derived in~\cite{Dumbser2018conformal} and for which strong hyperbolicity 
has been demonstrated for a particular choice of gauges. In \cite{Dumbser2020GLM}, the curl-free condition of certain evolution variables of the FO-CCZ4 system was taken into account via a novel hyperbolic generalized Lagrangian multiplier (GLM) \textit{curl-cleaning} technique.  

Numerical simulations of GRMHD and FO-CCZ4 at the aid of high order DG schemes with \textit{a posteriori} subcell finite volume limiter have been presented in~\cite{ADERGRMHD,Dumbser2018conformal,Dumbser2020GLM}, 
showing the appropriateness of the chosen governing PDE systems and the numerical methods employed for their solution. However, the obtained results also clearly indicate that with standard schemes for hyperbolic PDE one can reach only rather short simulation times (w.r.t astrophysical timescales)
and rather fine meshes (w.r.t astrophysical spatial scales) are required, even for the simulation of stationary equilibrium solutions of the governing PDE systems.
Thus, there is a necessity to endow existing methods with the ability of preserving additional 
physical properties of the underlying PDE system in such a way that spurious numerical modes can be eliminated,
numerical dissipation reduced and small-scale structures can manifest inside large-scale macroscopic models. In Newtonian computational fluid dynamics, such improved properties of a numerical scheme can be achieved at the aid of well balancing (WB), which allows to preserve certain stationary equilibrium solutions of the governing PDE system  \textit{exactly} at the discrete level. In the pioneering work of  Berm\'udez and V\'azquez in 1994 \cite{Bermudez1994} the authors have proposed an exactly well balanced Roe solver for lake at rest solutions of the shallow water equations. Further important results in this field have been achieved in \cite{bouchut2004nonlinear,audusse2004fast,rebollo2003family,Castro2008,gosse2001well,leveque1998balancing,perthame2001kinetic,tang2004gas,Castro2017Book}. More recent works are, for example, 
\cite{Gaburro2017CAFNonConf,Gaburro2018DiffuseInterface,chernyshenko2018hybrid,berberich2021high,castro2020well,arpaia2020well}.
Since it is impossible to mention all the relevant references here, we refer to the recent review paper~\cite{castro2020well}, which besides an interesting treatment of discontinuous equilibria contains a rather complete review of the wide literature on well balanced schemes.
Finally, we would like to underline that the development of WB and structure preserving schemes for astrophysical applications 
is a topic of growing interest, although the existing works so far only employ much simpler Newtonian PDE systems, see for example~\cite{BottaKlein,Kapelli2014,KM15_630,Klingenberg2015, schaal2015astrophysical,desveaux2016well,bermudez2016numerical,Gaburro2018MNRAS,klingenberg2019arbitrary,grosheintz2019high,thomann2019second} and the references therein, where well balanced schemes for the Euler equations of gasdynamics with Newtonian gravity were proposed.

\smallskip

The rest of the paper is organized as follows.
In Section~\ref{sec.Model} we introduce the governing PDE systems that are solved in this paper; 
in Section~\ref{sec.Method} we present our new well balanced second order finite volume scheme for numerical general relativity that allows to preserve smooth equilibria exactly at the discrete level. 
The increased stability and resolution gained with the use of the new scheme are shown via a series of numerical tests of increasing difficulty presented in Section~\ref{sec.Results}. We close the paper with some remarks and an outlook to future works in Section~\ref{sec.Conclusions}.

\section{Governing PDE systems}
\label{sec.Model}

All governing PDE systems for numerical general relativity used throughout this paper can be written as first order hyperbolic partial differential equations of the following general form: 
\be
\label{eq.generalform}
\Q_t + \nabla \cdot \F(\Q)  + \B(\Q) \cdot \nabla \Q = \mathbf{S}(\Q), \quad \x =(x_1, x_2,x_3) \in \Omega \subset \mathbb{R}^3,
\ee
where $\x=(x_1,x_2,x_3) = (x,y,z)$ is the spatial position vector, $t$ represents the time, $\Omega$ is the computational domain, 
$\Q = (q_1,q_2, \dots, q_{m})^T$ is the state vector defined in the space of the admissible states $\Omega_{\Q} \subset \mathbb{R}^{m}$, $ \F(\Q) = (\f(\Q), \g(\Q), \h(\Q))$ is the non linear flux tensor, $\B(\Q) = (\B_1(\Q), \B_2(\Q), \B_3(\Q))$ is a matrix collecting the nonconservative terms, and $\mathbf{S}(\Q)$ represents a nonlinear algebraic source term. 
The form~\eqref{eq.generalform} is said to be \textit{hyperbolic} if for all directions $\mathbf{n} \neq \mathbf{0}$ 
the matrix 
$
\mathbf{A}_n =  \left( \partial \mathbf{F} / \partial{\mathbf{Q}} + \mathbf{B}(\Q) \right) \cdot \mathbf{n}
\label{eqn.A.sys} 
$
has $m$ real eigenvalues and a full set of $m$ linearly independent eigenvectors. 

In this work we will consider problems that, due to their spherical symmetry, can be written as \textit{one-dimensional} hyperbolic PDE 
systems of the type
\be
\label{eq.generalform1D}
\Q_t + \f_{x}(\Q)  + \B_1(\Q) \cdot \Q_{x} = \mathbf{S}(\Q) - \SB(\Q,\Q_{y},\Q_{z}),
\ee 
where in $\SB$ we collect eventual non-zero terms depending on derivatives of the state variables with respect to the $y$ and $z$ coordinates. 
\blue{Throughout} this paper we employ the Einstein summation convention, which implies summation over two repeated indices and 
we use Greek letters for four-dimensional indices running from 0 to 3 and Latin letters for three-dimensional indices running from 1 to 3. 
We work in a geometrized set of units, in which the speed of light and the gravitational constant are set to unity, i.e. $c=G=1$.

The complete set of governing equations of the fully coupled FO-CCZ4 + GRMHD system reads as follows (see~\cite{Dumbser2018conformal,Dumbser2020GLM} and~\cite{ADERGRMHD}):   
\begin{subequations}
	\begin{eqnarray}
	\label{eqn.gamma}
	\partial_t\tilde\gamma_{ij}  
	&=&  {\beta^k 2 D_{kij} + \tilde\gamma_{ki} B_{j}^k  + \tilde\gamma_{kj} B_{i}^k - \frac{2}{3}\tilde\gamma_{ij} B_k^k }
	- 2\alpha \left( \tilde A_{ij} - {\frac{1}{3} \tilde \gamma_{ij} {\rm tr}{\tilde A} } \right),
	\\
	\label{eqn.alpha}
	{ \partial_t \ln{\alpha} }  &=&  { \beta^k A_k } - \alpha g(\alpha) ( K - K_0 - 2\Theta {c} ), 
	\qquad \qquad \partial_t K_0 = 0, \\
	\label{eqn.beta}
	\partial_t \beta^i   &=& 
	s \beta^k B_k^i +
	s f b^i, \\
	\label{eqn.phi}
	{ \partial_t \ln{\phi} }  &=&  { \beta^k P_k } + \frac{1}{3} \left( \alpha K - {B_k^k} \right),
	\end{eqnarray} 
\end{subequations}
\begin{subequations}
	\begin{eqnarray}
	\label{eqn.Aij}
	\partial_t\tilde A _{ij} - \beta^k \partial_k\tilde A_{ij}  + \phi^2 \bigg[ \nabla_i\nabla_j  \alpha - \alpha \left( R_{ij}+ \nabla_i Z_j + \nabla_j Z_i - 8 \pi S_{ij} \right) \bigg] 	&& \nonumber \\  
	-  \frac{1}{3} \tilde\gamma_{ij} \bigg[ \nabla^k \nabla_k \alpha - \alpha (R +  2 \nabla_k Z^k - 8 \pi S ) \bigg] &&    \nonumber \\	
	=  { \tilde A_{ki} B_j^k + \tilde A_{kj} B_i^k - \frac{2}{3}\tilde A_{ij} B_k^k }
	+ \alpha \tilde A_{ij}(K - 2 \Theta {c} ) - 2 \alpha\tilde A_{il} \tilde\gamma^{lm} \tilde A_{mj}, \ && 
	\\ \nonumber \\  
	\label{eqn.K}
	\partial_t K - \beta^k \partial_k K + \nabla^i \nabla_i \alpha - \alpha( R + 2 \nabla_i Z^i) 	&& \nonumber \\   
	=	\alpha K (K - 2\Theta {c} ) - 3\alpha\kappa_1(1+\kappa_2)\Theta + 4 \pi \alpha( S - 3 \tau), \ && 
	\\ \nonumber \\ 
	\label{eqn.theta}
\partial_t \Theta - \beta^k\partial_k\Theta - \frac{1}{2}\alpha {e^2} ( R + 2 \nabla_i Z^i) 	&& \nonumber \\   
=  \alpha {e^2} \left( \frac{1}{3} K^2 - \frac{1}{2} \tilde{A}_{ij} \tilde{A}^{ij}  - 8 \pi \tau \right) - \alpha \Theta K {c} - {Z^i \alpha A_i}
- \alpha\kappa_1(2+ \kappa_2)\Theta, \  &&  
	\\ \nonumber \\   	 
	\label{eqn.Ghat}
	\partial_t \hat\Gamma^i - \beta^k \partial_k \hat \Gamma^i + \frac{4}{3} \alpha \tilde{\gamma}^{ij} \partial_j K  
	- \tilde{\gamma}^{kl} \partial_{(k} B_{l)}^i
	- \tilde{\gamma}^{ik}  \left( \frac{1}{3}  \partial_{(k} B_{l)}^l 
	+ { s 2 \alpha   \tilde{\gamma}^{nm} \partial_k \tilde{A}_{nm}   }
	+ 2 \alpha  \partial_k \Theta \right) 
	&& \nonumber \\
	=  { \frac{2}{3} \tilde{\Gamma}^i B_k^k - \tilde{\Gamma}^k B_k^i  } +
	2 \alpha \left( \tilde{\Gamma}^i_{jk} \tilde{A}^{jk} - 3 \tilde{A}^{ij} P_j \right) -
	2 \alpha \tilde{\gamma}^{ki} \left( \Theta A_k + \frac{2}{3} K Z_k \right) -
	2 \alpha \tilde{A}^{ij} A_j && \nonumber \\
	- { 4 s \, \alpha \tilde{\gamma}^{ik} D_k^{~\,nm} \tilde{A}_{nm} } + 2\kappa_3 \left( \frac{2}{3} \tilde{\gamma}^{ij} Z_j B_k^k - \tilde{\gamma}^{jk} Z_j B_k^i \right) - 2 \alpha \kappa_1 \tilde{\gamma}^{ij} Z_j - 16 \pi \alpha \tilde{\gamma}^{ij} S_j,     &&
	\nonumber \\  
	&& \\ 
	\label{eqn.b}
	\partial_t b^i - s \beta^k \partial_k b^i = s \left(  \partial_t \hat\Gamma^i - \beta^k \partial_k \hat \Gamma^i - \eta b^i \right). && 
	\end{eqnarray} 
\end{subequations}
The use of the logarithms in the evolution equations for the lapse and the conformal factor is for convenience, in order to always guarantee \textit{positivity} of $\alpha$ and $\phi$ also at the 
\textit{discrete level}.  
In order to obtain a strongly hyperbolic first-order reduction, the following evolution system 
for the auxiliary variables 
$A_k := \partial_k \alpha / \alpha $, $B_k^{i} := \partial_k\beta^i$, 
$D_{kij} := \frac{1}{2}\partial_k\tilde\gamma_{ij}$ and 
$P_k :=  \partial_k \phi / \phi$ is added, accounting also for the stationary involutions of the governing PDE system:  
\begin{subequations}
	\begin{eqnarray}
	\label{eqn.A.GLM}
	\partial_t A_{k} - {\beta^l \partial_l A_k}  + \alpha g(\alpha) \left( \partial_k K - \partial _k K_0 - 2c \partial_k \Theta \right)
	+ {s \alpha g(\alpha) \tilde{\gamma}^{nm} \partial_k \tilde{A}_{nm} } 
	&& \nonumber \\   
	=  
	+ {2s\, \alpha g(\alpha) D_k^{~\,nm} \tilde{A}_{nm} }
	-\alpha A_k \left( K - K_0 - 2 \Theta c \right) h(\alpha) + B_k^l ~A_{l} \,, && \\  
	\partial_t B_k^i - s\beta^l \partial_l B_k^i   - s \left(  f \partial_k b^i -  \alpha^2 \mu \, \tilde{\gamma}^{ij} \left( \partial_k P_j - \partial_j P_k \right) \right) && \nonumber \\ 
		- s \left( \alpha^2 \mu \, \tilde{\gamma}^{ij} \tilde{\gamma}^{nl} \left( \partial_k D_{ljn} - \partial_l D_{kjn} \right)  \right)  
	= s B^l_k~B^i_l,  && \\ 
	\label{eqn.D.GLM}
	\partial_t D_{kij} - {\beta^l \partial_l D_{kij}} + s \left(
	- \frac{1}{2} \tilde{\gamma}_{mi} \partial_{(k} {B}_{j)}^m
	- \frac{1}{2} \tilde{\gamma}_{mj} \partial_{(k} {B}_{i)}^m
	+ \frac{1}{3} \tilde{\gamma}_{ij} \partial_{(k} {B}_{m)}^m  \right)  && \nonumber \\ 
	+  \alpha \partial_k \tilde{A}_{ij} 
	-   { \alpha \frac{1}{3} \tilde{\gamma}_{ij} \tilde{\gamma}^{nm} \partial_k \tilde{A}_{nm} }   
	=   B_k^l D_{lij} + B_j^l D_{kli} + B_i^l D_{klj} - \frac{2}{3} B_l^l D_{kij} 
		&& \nonumber  \\  
	-   \alpha \frac{2}{3}  \tilde{\gamma}_{ij} D_k^{~\,nm} \tilde{A}_{nm}  
	 - \alpha A_k \left( \tilde{A}_{ij} - \frac{1}{3} \tilde{\gamma}_{ij} {\rm tr} \tilde{A} \right), && \\
	\label{eqn.P.GLM}
	\partial_t P_{k} - \beta^l \partial_l P_{k}  - \frac{1}{3} \alpha \partial_k K
	+ \frac{1}{3} \partial_{(k} {B}_{i)}^i   - {s \frac{1}{3} \alpha \tilde{\gamma}^{nm} \partial_k \tilde{A}_{nm} } && \nonumber \\ 
	 = \frac{1}{3} \alpha A_k K + B_k^l P_l - {s \frac{2}{3} \alpha \, D_k^{~\,nm} \tilde{A}_{nm} }. 
	\end{eqnarray}
\end{subequations}
The matter and magnetic field evolution equations, which are  \textit{fully coupled} with the above FO-CCZ4 system read: 
\begin{eqnarray}
\label{eqn.mhd.mass}
\partial_t ( \sqrt{\gamma} D ) + \partial_i \left( \gamma^{\frac{1}{2}}\left(  \lapse v^i D -
\shift^i D \right) \right) & = & 0, \\ 
\partial_t ( \sqrt{\gamma} S_j ) + 
\partial_i \left( \gamma^{\frac{1}{2}}\left(  \lapse T^i_j - \shift^i S_j \right) \right)   && \nonumber \\   
= \gamma^{\frac{1}{2}} \left(  
\lapse T^{ik} C_{jik} + S_i B_j^i - \lapse \tau A_j \right),   \\ 
\partial_t ( \sqrt{\gamma} \tau ) + \partial_i \left( \gamma^{\frac{1}{2}}\left(  \lapse \left( S^i - v^i
D\right) - \shift^i \tau \right) \right) && \nonumber \\  
= \gamma^{\frac{1}{2}} \left( \lapse K_{ij} T^{ij}  - \lapse S^j  A_j  \right), \\ 
\partial_t ( \sqrt{\gamma} B^j ) + \partial_i \left( \gamma^{\frac{1}{2}}\left( \left( \lapse v^i - \shift^i \right) B^j
- \left( \lapse v^j - \shift^j \right) B^i + \varphi \delta^{ij} \right) \right) & = & 0, \\   
\label{eqn.mhd.phi}
\partial_t ( \sqrt{\gamma} \varphi ) + \partial_i \left( \gamma^{\frac{1}{2}}\left(   \lapse c_h^2 B^i - \shift^i \varphi \right) \right) & = & 0,  
\end{eqnarray}
with the abbreviation $C_{kij} = \frac{1}{\phi^2} \left( D_{kij} - P_k \tilde{\gamma}_{ij} \right)$ and 
where $T^{ij}$ denotes the spatial stress-energy tensor 
\begin{eqnarray}\label{eq:EMtensor}
T^{ij} &:= \rho h \Lor^2 {v^i v^j} - {E^i E^j} - {B^i B^j} + \left[ p +
\frac{1}{2} \left(E^2 + B^2\right) \right] \gamma^{ij} \nonumber \\ &=
{S^i v^j} + p_\text{tot} \gamma^{ij} - \frac{{B^i B^j}}{\Lor^2} - ({B}_k
{v}^k) {v^i B^j}\,,
\end{eqnarray}
with the enthalpy $\rho h = \rho + \frac{\gamma}{\gamma-1} p$, the total pressure defined as 
\begin{equation}
p_\text{tot}=p + p_\text{mag} = p + \frac{1}{2} \left[ B^2 / \Lor^2 + (B\cdot v
)^2 \right] \,.
\end{equation}
The governing PDE system~\eqref{eqn.gamma}--\eqref{eqn.P.GLM} contains the terms defined below:  
\begin{eqnarray}
{\rm tr} \tilde{A} & = &  \tilde{\gamma}^{ij} \tilde{A}_{ij}, \qquad \textnormal{ and } \qquad \tilde{\gamma} = \textnormal{det}( \tilde{\gamma}_{ij} ), \nonumber \\
\partial_k \tilde{\gamma}^{ij} & = &  - 2 \tilde{\gamma}^{in} \tilde{\gamma}^{mj} D_{knm} := -2 D_k^{~\,ij}, 
 \nonumber \\
\tilde{\Gamma}_{ij}^k &=& \tilde{\gamma}^{kl} \left( D_{ijl} + D_{jil} - D_{lij} \right), \nonumber \\
\label{eqn.dchr}
\partial_k \tilde{\Gamma}_{ij}^m & = & -2 D_k^{ml} \left( D_{ijl} + D_{jil} - D_{lij} \right)
+ \tilde{\gamma}^{ml} \left( \partial_{(k} {D}_{i)jl} + \partial_{(k} {D}_{j)il} - \partial_{(k} {D}_{l)ij} \right) , \nonumber \\ %
\Gamma_{ij}^k &=& \tilde{\gamma}^{kl} \left( D_{ijl} + D_{jil} - D_{lij} \right) - \tilde{\gamma}^{kl} \left( \tilde{\gamma}_{jl} P_i + \tilde{\gamma}_{il} P_j - \tilde{\gamma}_{ij} P_l \right) \nonumber \\ 
&=& \tilde{\Gamma}_{ij}^k - \tilde{\gamma}^{kl} \left( \tilde{\gamma}_{jl} P_i + \tilde{\gamma}_{il} P_j - \tilde{\gamma}_{ij} P_l \right),  \nonumber \\
\partial_k \Gamma_{ij}^m &=& -2 D_k^{ml} \left( D_{ijl} + D_{jil} - D_{lij} \right) +
2 D_k^{ml} \left( \tilde{\gamma}_{jl} P_i + \tilde{\gamma}_{il} P_j  - \tilde{\gamma}_{ij} P_l \right) \nonumber \\
&& 
- 2 \tilde{\gamma}^{ml} \left(  D_{kjl} P_i + D_{kil} P_j  - D_{kij} P_l \right)
\nonumber \\
&&	   + \tilde{\gamma}^{ml} \left( \partial_{(k} {D}_{i)jl} + \partial_{(k} {D}_{j)il} - \partial_{(k} {D}_{l)ij} \right)  
\nonumber \\ 
&& -
\tilde{\gamma}^{ml} \left( \tilde{\gamma}_{jl} \partial_{(k} {P}_{i)} + \tilde{\gamma}_{il} \partial_{(k} {P}_{j)}  - \tilde{\gamma}_{ij} \partial_{(k} {P}_{l)} \right) , \nonumber \\
R^m_{ikj} & = & \partial_k \Gamma^m_{ij} - \partial_j \Gamma^m_{ik} + \Gamma^l_{ij} \Gamma^m_{lk} - \Gamma^l_{ik} \Gamma^m_{lj}, \qquad 
R_{ij}  =  R^m_{imj}, \nonumber \\
\nabla_i \nabla_j \alpha &=& \alpha A_i A_j - \alpha \Gamma^k_{ij} A_k + \alpha \partial_{(i} {A}_{j)},  \qquad 
\nabla^i \nabla_i \alpha = \phi^2 \tilde{\gamma}^{ij} \left( \nabla_i \nabla_j \alpha \right), \nonumber \\
\tilde{\Gamma}^i & = &  \tilde{\gamma}^{jl} \tilde{\Gamma}^i_{jl},  \nonumber \\
\partial_k \tilde{\Gamma}^i & = & -2 D_k^{jl} \, \tilde{\Gamma}^i_{jl} + \tilde{\gamma}^{jl} \, \partial_k \tilde{\Gamma}^i_{jl}, \nonumber \\
Z_i &=& \frac{1}{2} \tilde\gamma_{ij} \left(\hat{\Gamma}^j-\tilde{\Gamma}^j \right), \qquad  Z^i = \frac{1}{2} \phi^2 (\hat\Gamma^i-\tilde\Gamma^i), \nonumber \\
\nabla_i Z_j &=& D_{ijl} \left(\hat{\Gamma}^l-\tilde{\Gamma}^l \right) + \frac{1}{2} \tilde\gamma_{jl} \left( \partial_i \hat{\Gamma}^l - \partial_i \tilde{\Gamma}^l \right) - \Gamma^l_{ij} Z_l, \nonumber \\
R + 2 \nabla_k Z^k & = & \phi^2 \tilde{\gamma}^{ij} \left( R_{ij} +
\nabla_i Z_j + \nabla_j Z_i \right)\,, \nonumber \\
h(\alpha) &=& \left( g(\alpha) + \alpha \frac{\partial g(\alpha)}{ \partial \alpha}  \right).
\nonumber 
%
%
%
%
\end{eqnarray}
The function $g(\alpha)$ in the PDE for the lapse $\alpha$ controls the slicing condition, where
$g(\alpha)=1$ leads to harmonic slicing and $g(\alpha)=2/\alpha$ leads to the so-called $1+\log$ 
slicing condition, see~\cite{Bona95b}.  
As already mentioned above, the auxiliary quantities $A_k$, $P_k$, $B^i_k$ and $D_{kij}$ are defined as (scaled) spatial gradients of the primary variables 
$\alpha$, $\phi$, $\beta^i$ and $\tilde{\gamma}_{ij}$, respectively, and read: 
\begin{equation}
\label{eq:Auxiliary}
A_i := \partial_i\ln\alpha = \frac{\partial_i \alpha }{\alpha}, \quad
B_k^{i} := \partial_k\beta^i,
\quad
D_{kij} := \frac{1}{2}\partial_k\tilde\gamma_{ij}, \quad
P_i       := \partial_i\ln\phi = \frac{\partial_i \phi}{\phi}.
\end{equation}
Hence, as a result, they must satisfy the following curl involutions or so-called second-order ordering constraints~\cite{Alic:2009,Gundlach:2005ta}:  
\begin{align}
\label{eqn.second.ord.const}
\mathcal{A}_{lk}   &:= \partial_l A_k     - \partial_k A_l     = 0, & 
\mathcal{P}_{lk}   &:= \partial_l P_k     - \partial_k P_l     = 0, \nonumber \\
\mathcal{B}_{lk}^i &:= \partial_l B_k^i   - \partial_k B_l^i   = 0, & 
\mathcal{D}_{lkij} &:= \partial_l D_{kij} - \partial_k D_{lij} = 0.   
\end{align}
In the governing PDE system above, we have already made use of these curl involutions by \textit{symmetrizing} 
the spatial derivatives of the auxiliary variables as follows:
\begin{equation}
\partial_{(k} {A}_{i)}     := \frac{ \partial_k A_i + \partial_i A_k       }{2}, \quad
\partial_{(k} {P}_{i)}     := \frac{ \partial_k P_i + \partial_i P_k       }{2}, \quad
\end{equation}
\begin{equation}
\partial_{(k} {B}^i_{j)}   := \frac{ \partial_k B^i_j + \partial_j B^i_k     }{2}, \quad
\partial_{(k} {D}_{l)ij}  := \frac{ \partial_k D_{lij} + \partial_l D_{kij} }{2}.
\label{eqn.symm.aux}
\end{equation}


\section{Numerical method}
\label{sec.Method}

In this section, we first briefly recall the standard (not well balanced) second order MUSCL-Hancock scheme applied to \eqref{eq.generalform1D} and then present the novel well balanced second order finite volume scheme. 
Throughout this paper, we assume that the equilibria to be preserved are smooth. In other words, equilibria with jumps at the element interfaces are not admitted. 

The one-dimensional computational domain $\Omega = [x_L, x_R]$ is covered with $N$ uniform segment elements of length $\Delta x=(x_R-x_L)/N$ denoted by $\Omega_i = [x_{i-1/2}, x_{i+1/2}]$, $i = 1, \dots, N$. 
The data at time $t^n$ are represented by the usual cell averages $\Q_i^n$
\be 
\label{eq.cellAverage} 
\Q_i^n = \frac{1}{\Delta x} \int_{\Omega_i} \Q(x,t^n)\,dx.
\ee  

\subsection{Standard second-order MUSCL-Hancock-type scheme} 

In the context of \textit{standard} MUSCL-Hancock-type  second order FV schemes \cite{toro-book}
for PDE \eqref{eq.generalform1D} the cell averages $\Q_i^n$ are evolved in time from $t^n$ up to $t^{n+1} = t^n + \Delta t$ through 
\be
\label{eq.StandardFV}
\Q_i^{n+1} = \Q_i^n - & \frac{\Delta t}{\Delta x}\left( \mathcal{F}_{i+\halb}\left (\q_{i+\halb}^{-}, \q_{i+\halb}^{+} \right ) 
- \mathcal{F}_{i-\halb}\left (\q_{i-\halb}^{-}, \q_{i-\halb}^{+} \right ) \right) \\[3pt] 
- & \frac{\Delta t}{\Delta x}\left( \mathcal{D}_{i-\halb}  \left (\q_{i-\halb}^{-}, \q_{i-\halb}^{+} \right ) 
+ \mathcal{D}_{i+\halb} \left (\q_{i+\halb}^{-}, \q_{i+\halb}^{+}\right ) \right) \\[3pt]
- & {\Delta t} \, \B\left (\q_i^{n+\halb} \right ) \cdot \partial_x \Q_{i}^n  +  {\Delta t} \, \mathbf{S} \left (\q_i^{n+\halb} \right ) - {\Delta t} \, \SB \left (\q_i^{n+\halb}, \partial_y \Q_i, \partial_z \Q_{ i} \right ).
\ee 

In~\eqref{eq.StandardFV}, $\mathcal{F}_{i\pm\halb}$ represent the numerical fluxes and the $\mathcal{D}_{i\pm\halb}$ are the nonconservative jump terms of~\eqref{eq.generalform} at the element interfaces given \blue{respectively} by 
\begin{equation} 
\label{eqn.fluxPC}
\mathcal{F}(\q^{-},\q^{+}) 
=\   \frac{1}{2} \left( \f(\q^{+}) + \f(\q^{-}) \right) 
-  \frac{1}{2} s_{\max} \left( \q^{+} - \q^{-} \right) 
\end{equation}
\blue{and}
\begin{equation} 
\mathcal{D}(\q^{-},\q^{+}) =  \frac{1}{2} \left(\int \limits_{0}^{1} {\mathbf{B}_1} \left(\mathbf{\Psi}(\q^{-},\q^{+},\tau) \right) \, ds \right)
\cdot\left(\q^{+} - \q^{-}\right)\!.
\end{equation}
In our scheme we employ a simple Rusanov-type Riemann solver, where  $s_{\max}$ is the maximum eigenvalue of the system matrices $\mathbf{A}(\q^{+})$ and $\mathbf{A}(\q^{-})$, and the theory introduced in~\cite{DalMaso1995} is used for integrating the nonconservative products along 
a Lipschitz continuous path $\mathbf{\Psi}(\q^-,\q^+, s)$ with $0 \leq s \leq 1$, $\mathbf{\Psi}(\q^-,\q^+, 0)
= \q^-$ and $\mathbf{\Psi}(\q^-,\q^+, 1) = \q^+$. Throughout this paper we employ the simple straight-line segment path
\be 
\label{eq.path_segments}
\mathbf{\Psi}(\q^-, \q^+, s) = \q^- + s \left( \q^+ - \q^- \right)\!, 
\qquad  s \in [0,1].
\ee

Moreover, in~\eqref{eq.StandardFV} the values 
\begin{equation} 
\q_{i-\frac{1}{2}}^{+} =\q_i^n(x_{i-\halb},t^{n+\halb}), \ \ \q_{i+\frac{1}{2}}^{-} =\q_i^n(x_{i+\halb},t^{n+\halb}), \ \
\q_{i}^{n+\halb} =\q_i^n(x_{i},t^{n+\halb}), 
\end{equation}
represent the evaluation of the reconstruction polynomials $\q_i^n(x,t)$ respectively at the two boundaries $x_{i\pm1/2}$ 
and in the barycenter $x_i$ of $\Omega_i$ at the time-midpoint of $[t^n, t^{n+1}]$ called $t^{n+1/2}$.
The reconstruction polynomial $\q_i^n(x,t)$ is computed through the MUSCL-Hancock strategy, see \cite{leer2,leer5,toro-book},  
in conjunction with the \textit{minmod} slope limiter. Thus we write it in the form of a linear polynomial in space and time that reads 
\be
\label{eq.predictor}
\q_i^n(x,t) = \w_i^n(x,t^n) + \partial_t\Q_i^n (t -t^n), \quad \text{with } \  \w_i^n(x,t^n) = \Q_i^n +  \partial_x \Q_{i}^n (x-x_i).
\ee 
In the above formula the spatial gradient is given by
\be
\label{eq.predictor_gradient}
\partial_x \Q_{i}^n = \text{minmod} \left ( \frac{\Delta \Q^n_{i+\halb}}{\Delta x},  \frac{\Delta \Q^n_{i-\halb}}{\Delta x} \right),
\ee 
with the jumps {$ \Delta \Q^n_{i-\halb} = \Q_i^n - \Q_{i-1}^n$} and  $ \Delta \Q^n_{i+\halb} = \Q_{i+1}^n - \Q_i^n $ across the left and the right cell boundary respectively, and the usual minmod function  
\be
\text{minmod}(a,b) = \begin{cases}
	0, \quad \text{if }  ab \le 0, \\
	a, \quad \text{if } |a| < |b|, \\
	b, \quad \text{if } |a| \ge |b|. 
\end{cases} 
\ee 
The term $\partial_t \Q_i^n$ indicates the approximation of the time derivative of $\Q$ and is computed using a discrete version of the governing equation
\begin{eqnarray}
\label{eq.predictor_timederivative}
 \partial_t \Q_i^n & = & - \frac{\f \left( \Q_i^n + \halb \Delta x \, \partial_x \Q_i^n \right) - \f\left( \Q_i^n - \halb \Delta x \, \partial_x \Q_i^n \right)}{\Delta x}  \nonumber \\ 
 && + \,\mathbf{S}(\Q_i^n) -\SB(\Q_i^n,\partial_y \Q_{i},\partial_z \Q_{i}) \! - \B_1(\Q_i^n)\cdot \partial_x \Q_{i}^n.
\end{eqnarray} 

Since we are working in the one-dimensional framework of \eqref{eq.generalform1D}, only the first component of the gradient of  $\Q$, i.e. $\partial_x \Q_i^n$ needs to be computed in each time step according to \eqref{eq.predictor_gradient}.
Instead, $\partial_y \Q_{i}$ and $\partial_z \Q_{i}$ are assumed to be known functions that are constant in time. Depending on the test problem, they may be non-zero. When not analytically available, they are computed once at the beginning of the simulation using a fourth order accurate finite difference approximation.
Finally, the time step $\Delta t$ is chosen according to a CFL condition 
\begin{equation}
\Delta t = \textnormal{CFL} \, \min \limits_{\Omega_i} \frac{\Delta x}{|\lambda_{\max,i}|}, \qquad \forall \Omega_i \in \Omega, 
\label{eq:timestep}
\end{equation}
where the Courant-Friedrichs-Levy number is set to $\textnormal{CFL}<1$ and $|\lambda_{\max,i}|$ is the maximum absolute value of the eigenvalues in the cell $\Omega_i$.

\subsection{Well balanced second order FV scheme}
\label{sec.MethodWB}

The scheme~\eqref{eq.StandardFV} and the above reconstruction procedure can be made well balanced by adopting 
the strategy first proposed in~\cite{Pares2006,Castro2006,Castro2008} and further developed for smooth equilibrium profiles in~\cite{berberich2021high,grosheintz2019high,castro2020well}.
The fundamental idea consists in exploiting the knowledge of the equilibrium profile $\Q^E(x)$ that needs to be preserved   in order to 
rewrite~\eqref{eq.StandardFV} as follows
\be
\label{eq.WBFV}
\!\!\!\!\!\!\!\!\!\!\!\!\!\!\!\!\!\!\!\!\!\!\!\!\!\!\!\!\!\Q_i^{n+1} = \Q_i^n  - & \frac{\Delta t}{\Delta x} \biggl [  \left( \mathcal{F}_{i+\halb}\left (\q_{i+\halb}^{-}, \q_{i+\halb}^{+} \right ) - \f(\Q_{i+\halb}^E) \right) 
+ \mathcal{D}_{i+\halb} \left (\q_{i+\halb}^{-}, \q_{i+\halb}^{+}\right )
 \\
   & \quad  -  \left( \mathcal{F}_{i-\halb}\left (\q_{i-\halb}^{-}, \q_{i-\halb}^{+} \right ) - \f(\Q_{i-\halb}^E) \right)  
+ \mathcal{D}_{i-\halb}  \left (\q_{i-\halb}^{-}, \q_{i-\halb}^{+} \right )  
  \biggr ]  \\[3pt]
- & {\Delta t} \left( \B\left (\q_i^{n+\halb} \right )  \partial_x \Q_{i}^n   - \B\left (\Q_i^{E} \right )  \partial_x \Q_{i}^E     \right) 
+ {\Delta t}  \biggl (  \mathbf{S} \left (\q_i^{n+\halb} \right ) - \mathbf{S} \left (\Q_i^{E} \right ) \biggr )\!\!\!\!\!\!\!\!\!\!\!\!\!\!\!\!\!\!\!\!\!\!\!\!\!\!\!\!\!\!\!\!\!\!\!\!\\
- &  {\Delta t} \, \biggl ( \SB \left (\q_i^{n+\halb}, \partial_y \Q_{i}, \partial_z \Q_{i} \right ) - \SB \left (\Q_i^{E}, \partial_y \Q_{i}^E, \partial_z \Q_{i}^E \right ) \biggr ).
\ee 
The \textit{validity } of~\eqref{eq.WBFV} can be easily \textit{justified} by first noticing that 
\begin{multline}\label{eq:equilibria}
\frac{\f(\Q^E_{i+\halb})-\f(\Q^E_{i-\halb})}{\Delta x}  +
\frac{1}{\Delta x} \int_{\Omega_i} \Bigl [ \B_1(\Q^E)\partial_x \Q^E +\mathbf{S}(\Q^E) \\
 -\SB \left(\Q^E,\partial_y \Q^E,\partial_z \Q^E\right) \Bigr] \, dx
\end{multline}
is \textit{exactly zero} by construction, since it coincides with the PDE system \eqref{eq.generalform1D} evaluated at the equilibrium.
Then, by subtracting \eqref{eq:equilibria} to the standard non well balanced FV solver \eqref{eq.StandardFV} and using the midpoint rule to approximate the integrals in \eqref{eq:equilibria}, we get exactly \eqref{eq.WBFV}. \textcolor{black}{As discussed later, for $\q_i(x,t)=\Q_i^E(x)$ one immediately obtains $\Q_i^{n+1}=\Q_i^n$ from \eqref{eq.WBFV}, independently of the quadrature rules employed in \eqref{eq:equilibria}, since all differences of the numerical solution with respect to the equilibrium solution in \eqref{eq.WBFV} cancel.}
For the sake of completeness, we recall that in \eqref{eq.WBFV}
\begin{equation}
\label{eq:eql1}
\Q^E_{i\pm\halb}=\Q^E(x_{i\pm \halb}), \quad \Q_i^E=\Q^E(x_i) 
\end{equation}
and
\begin{equation}
\label{eq:eql2}
\partial_x \Q^E_i =\frac{\Q^E_{i+\halb}-\Q^E_{i-\halb}}{\Delta x}, \ 
\partial_x \Q_i^n=\partial_x \Q^E_i + \partial_x \Q_i^{f,n},
\end{equation}
where $\Q_i^{f,n}$ is given in \eqref{eq:dfluc}. Finally, $\partial_y \Q^E_i$ and $\partial_z \Q^E_i$ are computed using the same procedure as the one employed for $\partial_y \Q_i$ and $\partial_z \Q_i$. 

A similar scheme could be derived following the strategy described in  in~\cite{Gaburro2018MNRAS}, where a special path is defined to achieve the well balanced property,  indeed in that case the proposed path was composed of two components 
\be
\label{eq.path_sumEqFluct}
\Phi(s,\q^-,\q^+)=\Phi^E(s,\q_E^-, \q_E^+)+\Phi^f(s,\q_f^-, \q_f^+), 
\ee
where $\Phi^E(s,\q_E^-, \q_E^+)$ is a re-parametrization of the equilibrium solution of interest that connects the two equilibrium states $\q_E^-$ and $\q_E^+$ with each other, and  $\Phi^f(s,\q_f^-, \q_f^+)=  \q_f^- + s(\q_f^+ - \q_f^-)$ is a standard segment path but acting \textit{only} on the {\it fluctuations} with respect to $\q_E$, defined as $\q_f^- = \q^- - \q_E^-$ and $\q_f^+ = \q^+ - \q_E^+$, respectively.
Here, the fact of computing separately the contributions coming from the equilibrium components of the solution and subtracting them pointwise allows to immediately cancel the numerical errors introduced in that part of the computation and allows an increased resolution on the fluctuations with respect to 
that equilibrium. 

Also the reconstruction procedure~\eqref{eq.predictor}
is modified according to the strategy proposed \blue{in} \cite{Castro2008} that consists in reconstructing only the fluctuations w.r.t the equilibrium and summing up the exact equilibrium value. This gives 
\be
\label{eq.WBpredictor}
&\q_i^n(x,t) = \Q_i^E(x) +  \w_i^{f,n}(x,t) + \partial_t\Q_i^{f,n} (t -t^n), \ \text{ with }\\
&\w_i^{f,n}(x,t) = \Q_i^{f,n} +  \partial_x \Q_{i}^{f,n} (x-x_i),
\ee 
where $\w_i^{f,n}$ and $\partial_t\Q_i^{f,n}$ are respectively obtained with~\eqref{eq.predictor_gradient} and~\eqref{eq.predictor_timederivative},
but applied to the fluctuations $\Q_i^{f,n} = \Q_i^n - \Q^E_i$ around the equilibrium, i.e. 
\begin{equation}
\label{eq:dfluc}
\partial_x \Q_{i}^{f,n} = \text{minmod} \left ( 
\frac{ \Q^{f,n}_{i+1}-\Q^{f,n}_{i}}{\Delta x},  
\frac{ \Q^{f,n}_{i}-\Q^{f,n}_{i-1}}{\Delta x} 
\right);
\end{equation}
\blue{finally, for what concerns the time evolution $\partial_t\Q_i^{f,n}$, it is again recovered from a discrete version of the governing PDE \eqref{eq.generalform1D} evaluated at the reconstructed values minus the same PDE evaluated at the equilibrium, i.e. 
\be
	\label{eq.predictor_timederivativeWB}
	\partial_t\Q_i^{f,n} = & - \frac{1}{\Delta x}  \Biggl[ \ \, \Biggl( \f \left( \Q_i^E(x_{i+\halb} ) + \w_i^{f,n}(x_{i+\halb},t) \right) - \f \left( \Q_i^E(x_{i+\halb} ) \right) \Biggr)   \\
	& \qquad\quad  -  \Biggl( \f\left( \Q_i^E(x_{i-\halb} ) + \w_i^{f,n}(x_{i-\halb},t) \right) - \f\left( \Q_i^E(x_{i-\halb}) \right)    \Biggr) \Biggr ]  \\[2pt]
	& + \,\mathbf{S}(\Q_i^n)-\mathbf{S}(\Q_i^E) -\SB(\Q_i^n,\partial_y \Q_{i},\partial_z \Q_{i}) +\SB(\Q_i^E,\partial_y \Q_{i}^E,\partial_z \Q_{i}^e) \\[2pt]
	& - \B_1(\Q_i^n)\cdot \partial_x \Q_{i}^n + \B_1(\Q_i^E)\cdot \partial_x \Q_{i}^E.  
\ee 
}

Finally, it is trivial to \textit{prove} that the scheme~\eqref{eq.WBFV} associated with the reconstruction procedure~\eqref{eq.WBpredictor} with the definitions \eqref{eq:eql1} and \eqref{eq:eql2},
is well balanced for any smooth equilibrium $\Q^E(x)$. Indeed, if the initial condition coincides with the chosen equilibrium, i.e $\Q_i^0=\Q^E(x_i)$, then at each timestep the fluctuations will be automatically zero and the reconstruction polynomial will be set equal to the equilibrium $\q(x,t)=\Q^E(x)$, according to~\eqref{eq.WBpredictor}. At this point each term of~\eqref{eq.WBFV} cancels exactly, and $\Q_i^{n+1}=\Q_i^{n}=\Q_i^E$ for any cell $\Omega_i$ of the domain.
It should be underlined that this proof is particularly simple due to the fact that the equilibria under consideration are smooth, $\q_{i\pm\halb}^{-}= \q_{i\pm\halb}^{+}=\Q^E_{i\pm\halb}$, 
and the numerical flux  $\mathcal{F}_{i\pm \halb}\left (\q_{i\pm\halb}^{-}, \q_{i\pm\halb}^{+} \right )=\f(\Q^E_{i\pm \halb})$.
We furthermore remark that the well balanced scheme~\eqref{eq.WBFV} will behave as \blue{the} standard scheme~\eqref{eq.StandardFV} when $\Q_i^n$ is far away from the chosen equilibria.


\section{Numerical results}
\label{sec.Results}

In this section, we show the increased resolution and stability of our second order well balanced finite volume method~\eqref{eq.WBFV}
with respect to a standard not well balanced second order scheme~\eqref{eq.StandardFV} for two typical applications in general relativity,
namely the Michel-Bondi accretion disk (Section~\ref{ssec.AccretionDisk}) and the TOV neutron star (Section~\ref{ssec.TOVStar}).
The considered problems are all characterized by a spherical symmetry, so in order to ease the comprehension we adopt the following notation $\x = (x_1,x_2,x_3) = (r, \theta,\phi)$.

We proceed as follows: 
after verifying that for an equilibrium initial condition $\Q(\x,0)=\Q^E(\x)$ the new well balanced scheme is always exact up to machine precision,  
we test the performance of the well balanced algorithm and a classical not well balanced method on initial conditions that could be considered as perturbations of the chosen equilibrium. 
As expected, for large perturbations the two schemes behave practically the same,
while for small perturbations (which are indeed the target of this work) 
the WB method is more accurate and more stable, hence allowing much longer simulations compared to a standard scheme.  
In particular, for what concerns the TOV star, 
we first study 
i) the \textit{evolution of the matter} via the GRMHD system when subject to a pressure perturbation on a \textit{fixed} background space--time \textit{metric} (Cowling approximation, see Section~\ref{ssec.TOVGRMHD}), then
ii) the \textit{evolution} of the space--time \textit{metric}, subject to a perturbation of the extrinsic curvature, but assuming the \textit{matter} variables as externally given and \textit{fixed} (anti-Cowling approximation, see Section~\ref{ssec.TOVanticowling}), and finally 
iii) the \textit{fully coupled evolution of matter and metric} when subject to either a pressure or a space--time perturbation (see Section~\ref{ssec.TOVfullycoupled}).

\blue{Finally, for the sake of completeness, we also verify numerically the order of convergence of the employed well balanced scheme, see Section~\ref{ssec.convergence}.}

\subsection{Michel accretion disk}
\label{ssec.AccretionDisk}

\begin{figure}
	\centering
	\includegraphics[width=0.33\linewidth]{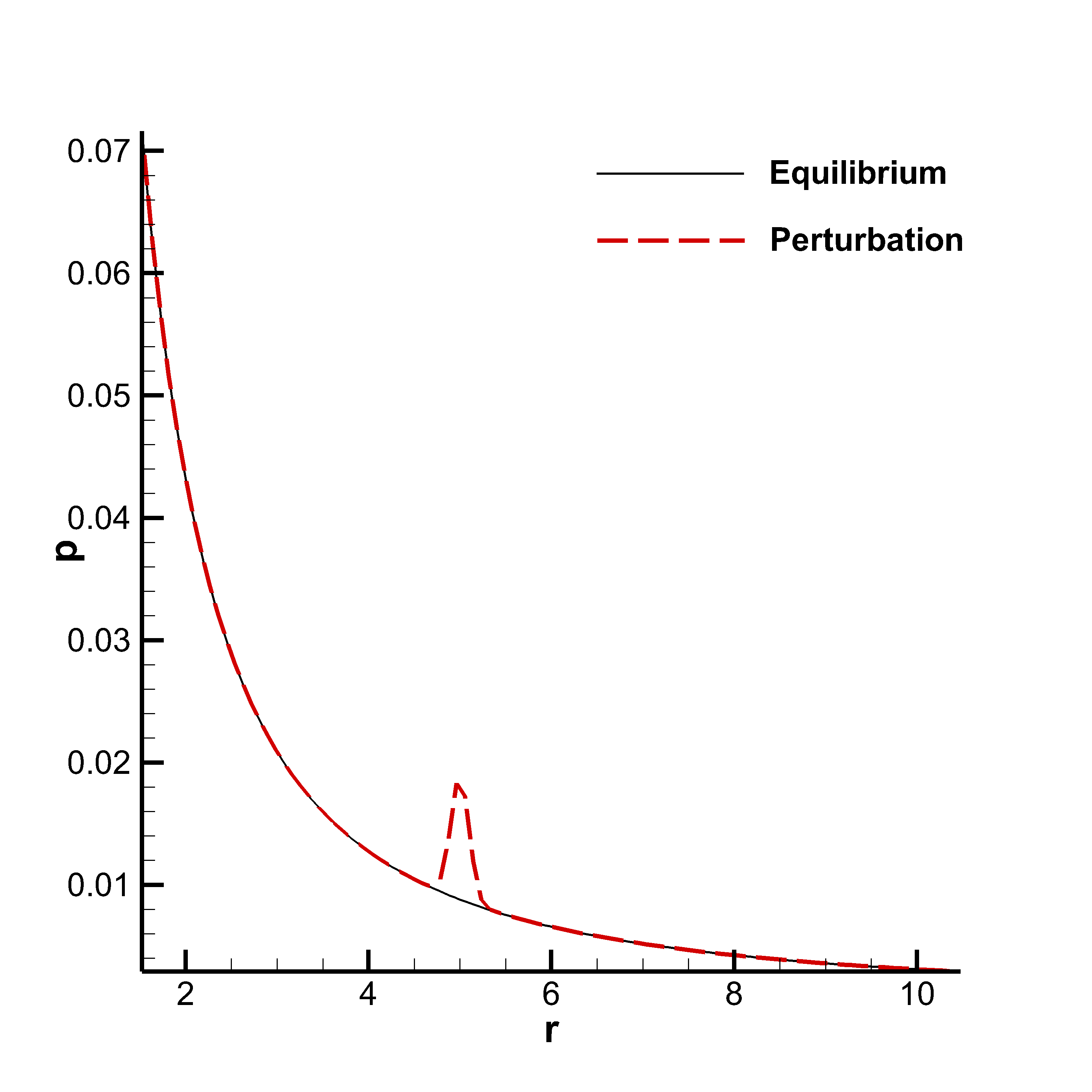}%
	\includegraphics[width=0.33\linewidth]{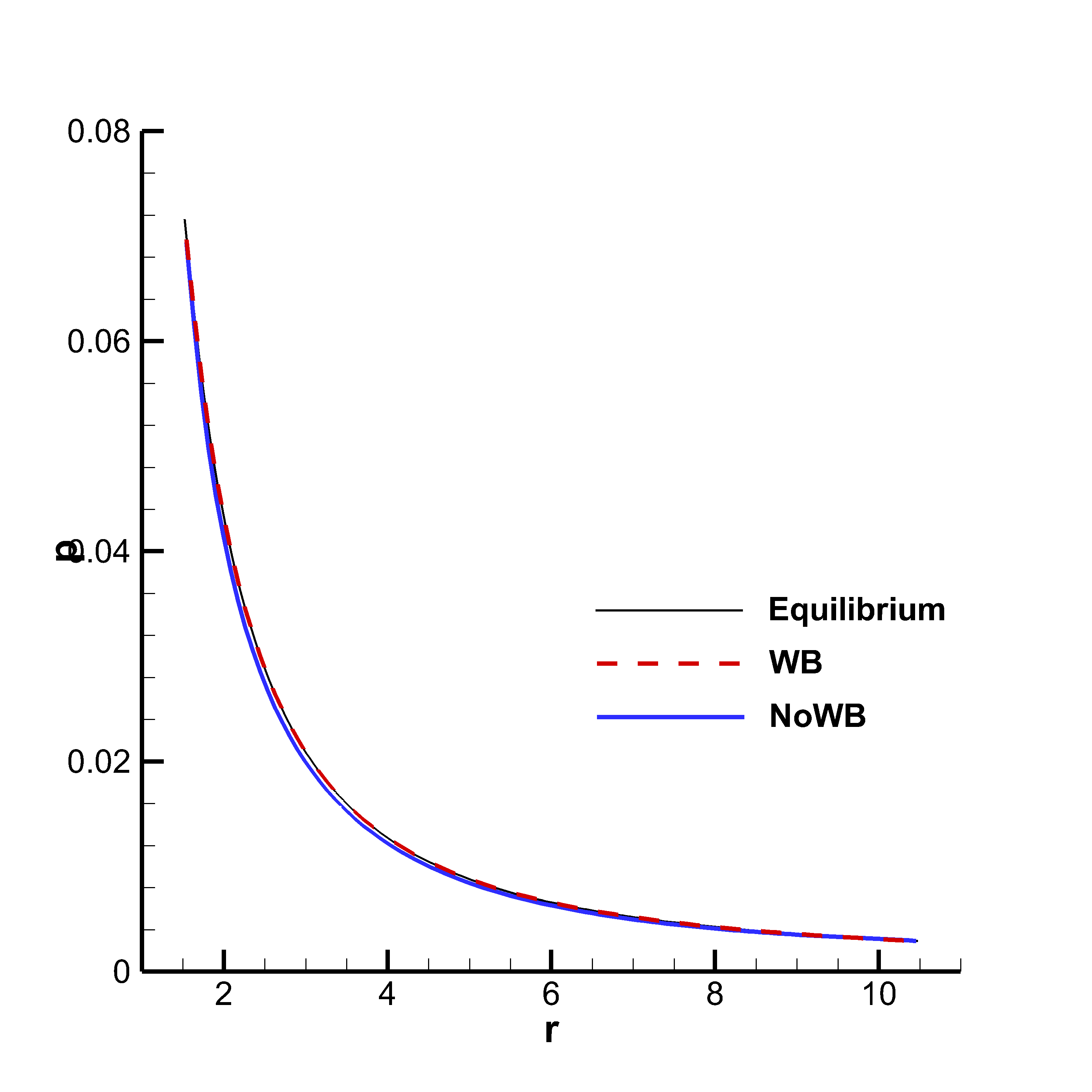}%
	\includegraphics[width=0.33\linewidth]{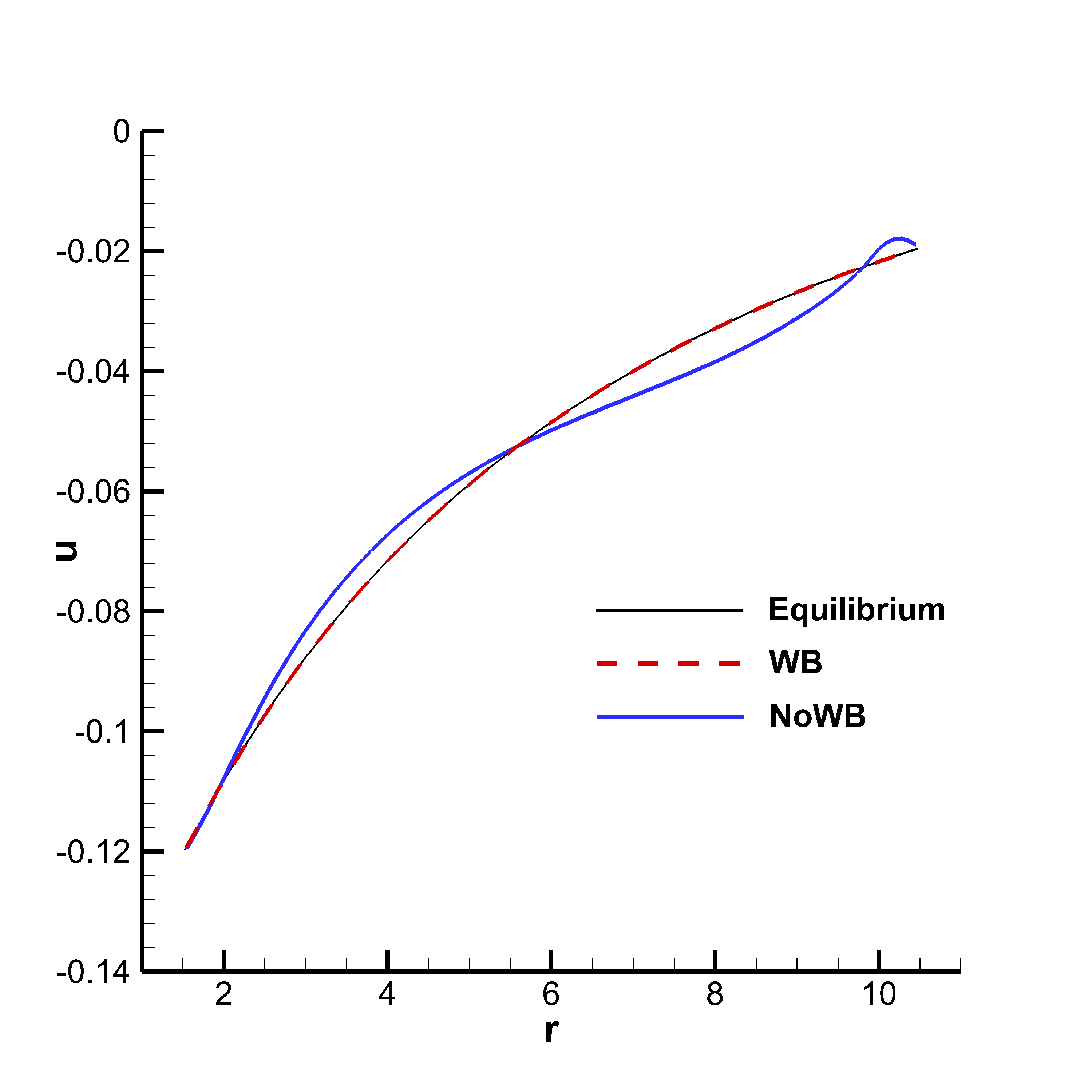}
		\vspace{-7pt}
	\caption{Michel accretion disk, equilibrium solution of GRMHD. 
		We show the equilibrium profile (continuous black line) for pressure (left, middle) and velocity (right), and the initial perturbed pressure condition (red dashed line, left).
		Then, in the middle and right panels we report the numerical results obtained at time $t_f = 10^5$ with the well balanced scheme (dashed red lines) and the not well balanced scheme (continuous blue lines). We remark that the initial bump is expected to split into two waves traveling out of the domain so that after a long simulation time the equilibrium profile should be recovered: this is the case with the WB scheme while the not WB scheme destroys the equilibrium profile (note in particular the completely unstable velocity profile, right panel).
	}
	\label{fig.grmhd_disk}
\end{figure}

As first test case we consider a smooth flow with an analytical solution on a curved space--time:  
it consists in the stationary spherical transonic accretion of an isentropic fluid onto a
non-rotating black hole and it is known as \textit{Michel-Bondi} solution~\cite{michel1972accretion}. 
The explicit expressions of the lapse, the shift and
the spatial metric of a Kerr black hole with mass $M$ and spin $a$ 
in the employed spherical Kerr-Schild (KS) coordinates
$(r,\theta,\phi)$ are given by~\cite{Komissarov04}
\be
& \alpha = (1+z)^{-\frac{1}{2}}, \quad \beta^i = \left( \frac{z}{1+z},0,0
\right), \\[3pt]
& \gamma_{ij} = \left( \begin{array}{ccc} 1 + z & 0 & - a \sin^2
\theta (1 + z) \\ 0 & \rho^2 & 0 \\ - a \sin^2 \theta (1 + z) &
0 & {\Sigma} \sin^2 \theta/{\rho^2}
\end{array} \right),
\label{eq.kerr.metric.spherical}
\ee
with
\be
& \rho^2 = r^2 + a^2 \cos^2 \theta, \qquad z = \frac{2 r}{\rho^2}, \\
&\Delta = r^2 + a^2 - 2 M r, \qquad \Sigma = (r^2 + a^2)^2 - a^2
\Delta \sin^2 \theta.
\ee
By considering this metric with $a=0$ and by fixing the mass $M=1$, the critical radius $r_c=8\,M$ and the critical
density $\rho_c M^2=1/16$, the Michel equilibrium solution $\Q^E(r)$ can be determined analytically, see for example~\cite{Rezzolla_book:2013}. This test is performed with the so-called Cowling approximation, i.e. the background space-time is assumed to be fixed. 
We perform this test on a spatial $1$D domain $\Omega$ with $r \in [1.5,10.5]$, $\theta =\pi/2$ and $\phi=0$, discretized with $N=100$ intervals and evolved up to the final time $t_f = 10^5$.
We have first verified that for an initial condition equal to the equilibrium $\Q^E(r)$ the WB scheme does not exhibit any numerical errors, and then we have perturbed the initial condition by adding a Gaussian bump to the equilibrium pressure profile as 
\be
p = p^E + 10^{-3} \exp \left ( -0.5 \, \frac{(r-5)^2}{0.5^2} \right )
\ee 
(see the left image of Figure~\ref{fig.grmhd_disk}).
The numerical results obtained with the new WB scheme and with a standard second order not well balanced scheme are reported in Figure~\ref{fig.grmhd_disk} and show how the WB scheme is able to recover the equilibrium profile once the bump has exited the domain, while this is not the case for the not WB scheme.

\subsection{Solution of a Riemann problem with prescribed Michel equilibrium}
\label{ssec.RP}

\begin{figure}
	\centering
	\includegraphics[width=0.45\linewidth]{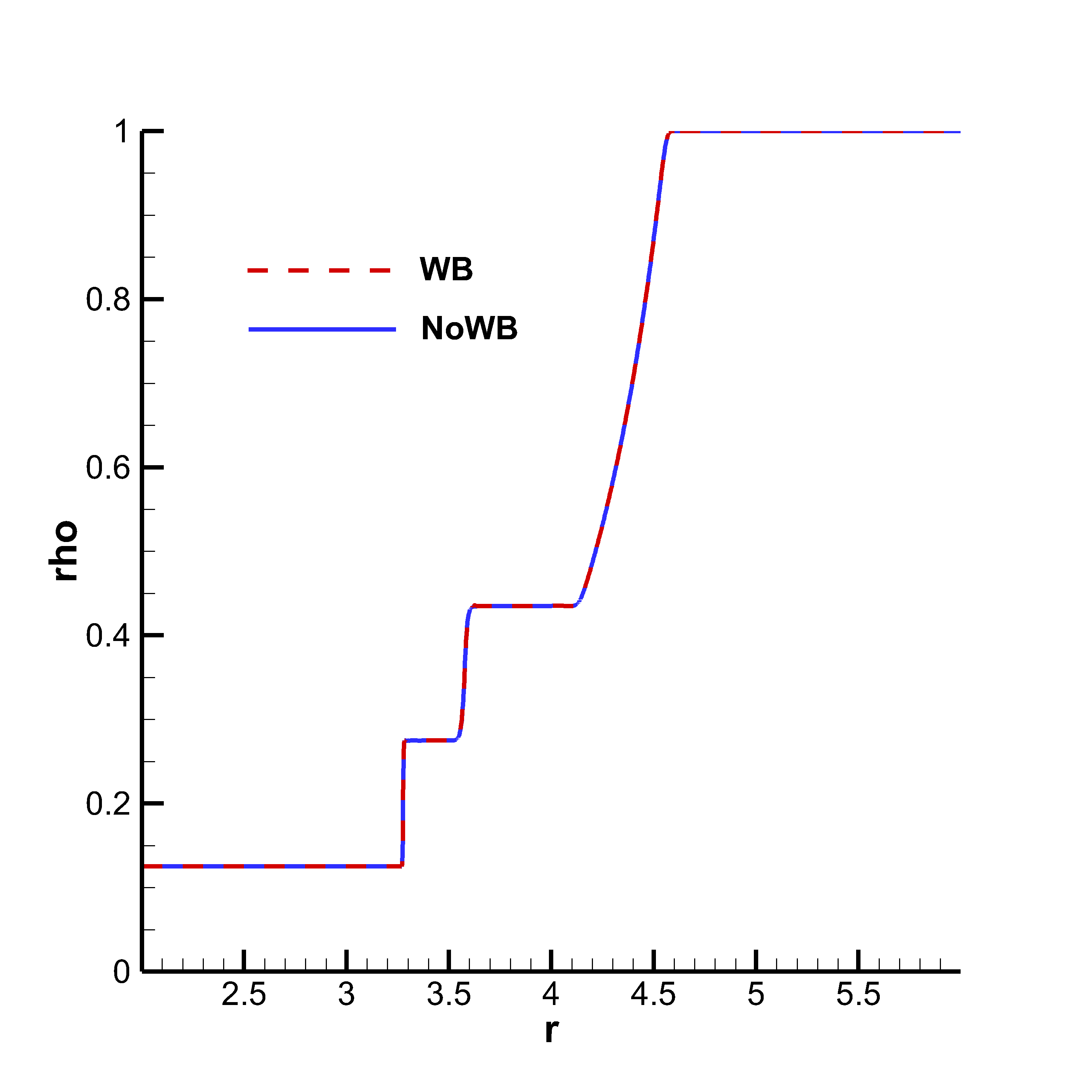}
	\includegraphics[width=0.45\linewidth]{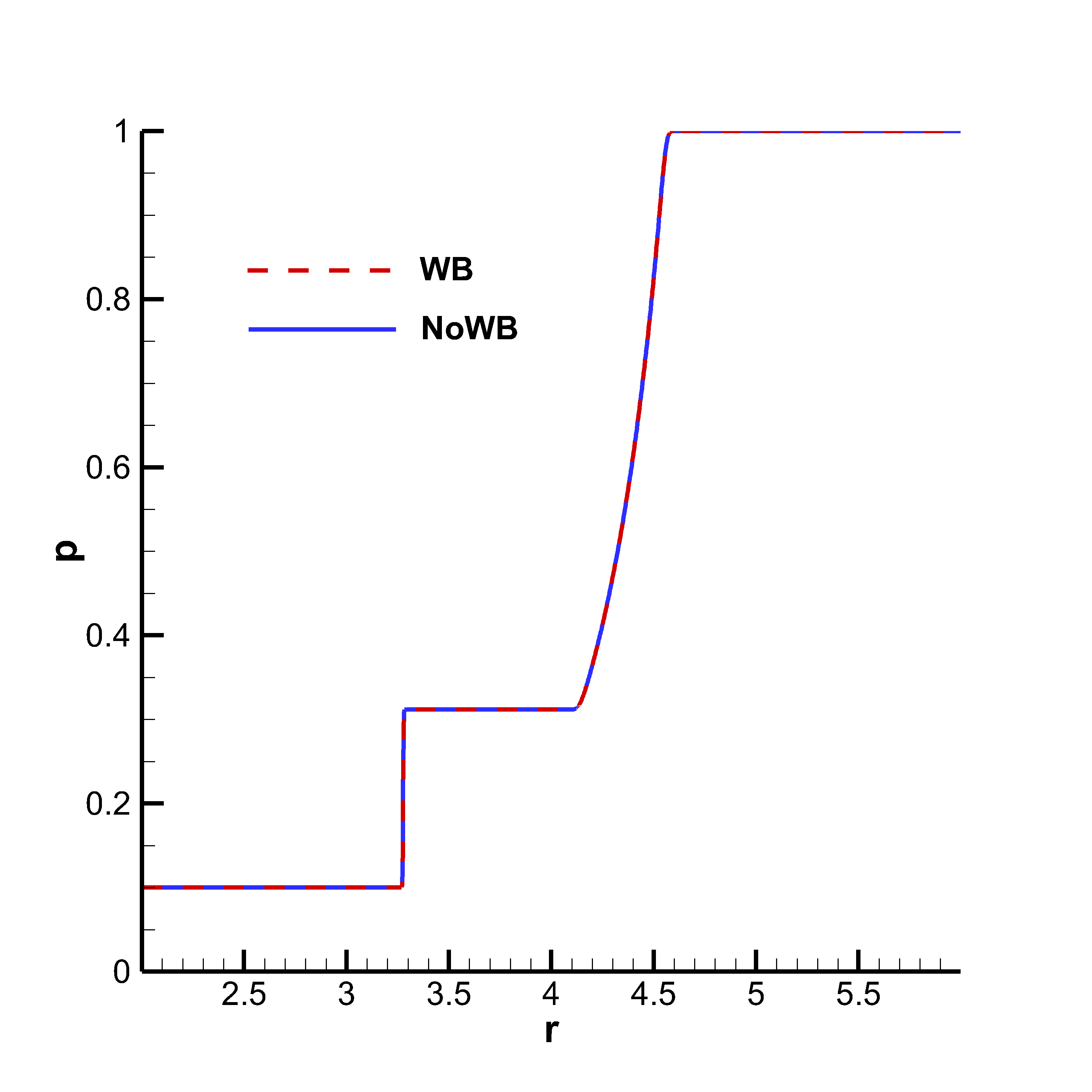}
	\vspace{-10pt}
	\caption{GRMHD Riemann problem. We show the numerical results for the density (left) and the pressure (right) obtained with a standard not WB scheme (blue) which coincide with those obtained with the new WB scheme (red) although the latter is set to maintain the Michel equilibrium.  }
	\label{fig.grmhdRP}
\end{figure}

We choose the second test case in order to show numerically that the WB property of the scheme, which radically improves its resolution near the equilibria, does not affect negatively any capability of the underlying second order scheme when performing simulations far away from the chosen equilibrium.
Thus, while choosing as equilibrium the Michel solution over the KS coordinates described in the previous Section~\ref{ssec.AccretionDisk}, we use our schemes to solve the following Riemann problem~\cite{ADERGRMHD}
\be
(\rho, \mathbf{u}, p, \mathbf{B}) =
\begin{cases}
(0.125, \mathbf{0}, 0.1, mathbf{0}) \quad \text{ if }  r \le 4  \\
(1, \mathbf{0}, 1, mathbf{0}) \quad \text{ if }  r > 4  
\end{cases}
\ee
over a flat Minkowski space--time, that is $\alpha = 1$, $\beta^i = (0,0,0)$ and $\gamma_{ij} = \mathbf{I}$, on a spatial $1$D domain $\Omega$ with $x_1 \in [2,6]$.
Note that, even if the initial condition completely differs both for matter and metric from the chosen equilibrium, the WB scheme is able to return the same results of the not WB scheme, see Figure~\ref{fig.grmhdRP}. Indeed far from the equilibrium the two schemes are both nominally second order accurate shock capturing TVD schemes and are expected to return equally accurate results.

\subsection{TOV Neutron star}
\label{ssec.TOVStar}

The next phenomenon that we want to investigate is the 
long time evolution of a stable Tolman-Oppenheimer-Volkoff (TOV) neutron star and the effects 
of small perturbations of metric and matter variables on the entire system.
For the details on the derivation of the radially symmetric TOV solution, 
we refer to~\cite{Tolman,Oppenheimer39b,Wald84,Carroll2003,bugner}. 

In the following, we will consider three different situations, 
namely the matter evolution on a fixed space--time metric through the GRMHD model (Cowling approximation), 
the space--time metric evolution in the anti-Cowling approximation through the FO-CCZ4 model, 
and the fully coupled evolution of metric and matter through the combined  FO-CCZ4+GRMHD model.
In the three cases we run the test problem on a $1$D domain $\Omega$ with $r\in[0.5, 14]$, $\theta=\pi/2$, $\phi=0$ discretized with $N=1000$ intervals.

To impose the \textit{boundary conditions} we proceed as follows. 
First, we use one ghost cell at each boundary set on the equilibrium value $\Q^E$. 
Next, we  define a left and a right sponge layer $\mathcal{S_{L,R}}$ (see also~\cite{BassiBusto}) so that 
\be
\label{eq.Pert_P}
x \in \mathcal{S_{L,R}} \quad \text{ if } \ |x-x_{L,R}| < s_\mathcal{L,R}
\ee 
on which the numerical solution is redefined as 
\be
\tilde{\q} = (1-\epsilon_{L,R})\q_h + \epsilon_{L,R}\Q^E.
\ee 
\blue{These} conditions are employed to impose an absorbing boundary condition which reduces the reflection of waves, 
and allows in particular the numerical errors traveling through the cleaning variables (as $\theta$ in the FO-CCZ4 model) to exit the domain.

To check the \textit{validity} of the obtained numerical results we will monitor 
i) the profile of the solution,
ii) the oscillation of the mass density at the center of the star, and
iii) the temporal evolution of the {Hamiltonian} and {momentum} constraints, which are 
nonlinear elliptic involutions (hence not containing time derivatives) that should be satisfied everywhere
\be
& \mathcal{H} = R_{ij} g^{ij} - K_{ij} K^{ij} + K^2 - 16 \pi \tau = 0, \\
& \mathcal{M}_i = \gamma^{jl} \left( \partial_l K_{ij} - \partial_i K_{jl} - \Gamma^m_{\,j,l} K_{mi} + \Gamma^m_{\,ji} K_{ml} \right) - 8 \pi S_i  = 0.
\label{eq.constraints} 
\ee
At this point, we would like to remark that the TOV equilibrium solution is not available in the form of an analytical expression, 
but just as the numerical solution of a system of four ODEs (see~\cite{bugner} for example). 
Because of that, already the initial equilibrium condition does not satisfy exactly~\eqref{eq.constraints}. 
Our aim will be thus preserving constantly the initial values of $\mathcal{H}$ and $\mathcal{M}_i$.

Finally, we emphasize that when the imposed initial conditions coincide exactly with the TOV equilibrium, 
the WB scheme is able to preserve it \textit{indefinitely} without the introduction of any numerical errors and to perfectly maintain constants the \blue{values of the constraints}. \blue{This well balanced property of our algorithm is further numerically verified in Section \ref{ssec.WBtest}.}

\blue{
	\subsubsection{Numerical proof of the well balanced property of our algorithm}
	\label{ssec.WBtest}
	First of all, in order to present a numerical proof of the well balanced capabilities of the proposed second order well balanced scheme we have set up the following test.
	
	Usually one claims that this property is verified when an equilibrium initial condition is preserved with machine accuracy for very long computational times.
	For our well balanced scheme this is trivially verified for any smooth equilibrium initial conditions $\Q(x,0)$ taken equal to the equilibrium profile $\Q^E$
	to be preserved. Since no numerical errors are introduced by the scheme, this property can be easily verified also numerically.
	
	To make this test more challenging and convincing, we have added a small random perturbation of the order of machine accuracy
	to the initial equilibrium profiles and we have monitored its evolution working in \textit{single}, \textit{double} and \textit{quadruple} precision. 
	We have applied this approach for all the equilibrium profiles presented in this paper.	
	In particular, we have modified the density and the pressure when employing the GRMHD equations, and the trace of the extrinsic curvature $K$ when using the FO-CCZ4 system or the fully coupled approach, with random perturbations of the order of the corresponding machine precision used for the simulations. In Table \ref{tab.WBmachineprecision} we have reported the numerical errors with respect to the equilibrium profiles for several variables of interest at short and long simulation times. In all the cases the numerical errors remain of the order of machine precision.

	\begin{landscape} 
		\begin{table*}\blue{ 
				\caption{Numerical proof of the well balanced property of the new well balanced finite volume scheme presented in this paper, as described in Section \ref{ssec.WBtest}.} 
				\label{tab.WBmachineprecision}
				\begin{center} 	
					\begin{tabular}{|c|cccc|cccc|cccc|} 
						\hline 
						Precision                   &         \mcolN{4}{Single}                           &         \mcolN{4}{Double}                                 &     \mcolN{4}{Quadruple}                         \\[1pt]
						Time                        & $\Delta t$   &   $1$      &   $100$   &   $1000$    & $\Delta t$   &   $1$      &   $100$   &   $1000$     &$ \Delta t$   &  $1$         &   $100$   &   $1000$    \\[1pt] 
						\hline 			                                                                  
						{\small Michel }            &              &            &           &             &              &            &           &              &              &              &           &             \\[1pt]
						$\rho$                      &   \!1E-07    &  8E-08     &  7E-08    &  9E-08\!    &   \!2E-14    &  2E-14     &  4E-14    &  4E-14\!     &  \!1E-28     &   1E-28      &  5E-29    &  1E-31\!    \\[1pt]
						$ p $                       &   \!2E-07    &  2E-07     &  2E-07    &  2E-07\!    &   \!1E-14    &  1E-14     &  1E-14    &  1E-14\!     &  \!1E-28     &   2E-28      &  1E-29    &  1E-32\!    \\[1pt]
						\hline 			                                                                                                                                                                                 
						{\small TOV GRMHD}          &              &            &           &             &              &            &           &              &              &              &           &             \\[1pt]
						$\rho$                      &   \!2E-07    &  3E-06     &  1E-05    &  1E-05\!    &   \!2E-14    &  7E-13     &  3E-12    &  2E-12\!     &  \!2E-28     &   7E-27      &  3E-26    &   2E-26\!   \\[1pt]
						$ p $		                &   \!2E-07    &  2E-07     &  5E-07    &  2E-07\!    &   \!2E-14    &  2E-14     &  4E-14    &  5E-14\!     &  \!2E-28     &   2E-28      &  4E-28    &   3E-28\!   \\[1pt]
						\hline 			                                                                                                                                                                                 
						{\small TOV anti-Cowling}   &              &            &           &             &              &            &           &              &              &              &           &             \\[1pt]         
						$K$					        &   \!2E-07    &  2E-07     &  3E-07    &  5E-07\!    &   \!2E-14    &  2E-14     &  5E-14    &  7E-14\!     &  \!2E-28     &   2E-28      &  2E-28    &   6E-28     \\[1pt]
						$\theta$	    	        &   \!0        &  5E-09     &  7E-09    &  7E-09\!    &   \!4E-19    &  6E-15     &  1E-14    &  4E-14\!     &  \!2E-33     &   4E-30      &  2E-28    &   2E-27          \\[1pt]
						\hline 			                                                                                                                                                         
						{\small TOV fully coupled}  &              &            &           &             &              &            &           &              &              &              &           &             \\[1pt]
						$\rho$                      &     0        &   1E-14    &  1E-07    &  9E-06\!    &     0        &   1E-19    &  4E-15    &  5E-14\!     &   0          &   2E-31      &  1E-29    &  4E-30\!    \\[1pt]
						$K$                         &   \!2E-07    &   2E-07    &  8E-06    &  9E-06\!    &   \!2E-14    &   2E-14    &  8E-15    &  4E-14\!     &   \!2E-28    &   2E-28      &  1E-28    &  9E-30\!    \\[1pt]
						$\theta$                    &   \!1E-11    &   5E-08    &  5E-06    &  9E-06\!    &   \!3E-19    &   2E-15    &  9E-15    &  3E-14\!     &   \!1E-32    &   1E-30      &  4E-29    &  7E-30\!    \\[1pt]
						$\gamma_{11}$               &     0        &   0        &  1E-05    &  2E-05\!    &   \!2E-18    &   2E-15    &  5E-15    &  3E-14\!     &   0          &   1E-31      &  1E-29    &  2E-29\!    \\[1pt]
						\hline 		                                                                                     		
					\end{tabular}		
			\end{center}  }
		\end{table*}
	\end{landscape} 
}

\subsubsection{TOV neutron star simulated with the GRMHD system in Cowling approximation}
\label{ssec.TOVGRMHD}

\begin{figure}
	\centering
	\includegraphics[width=0.45\linewidth]{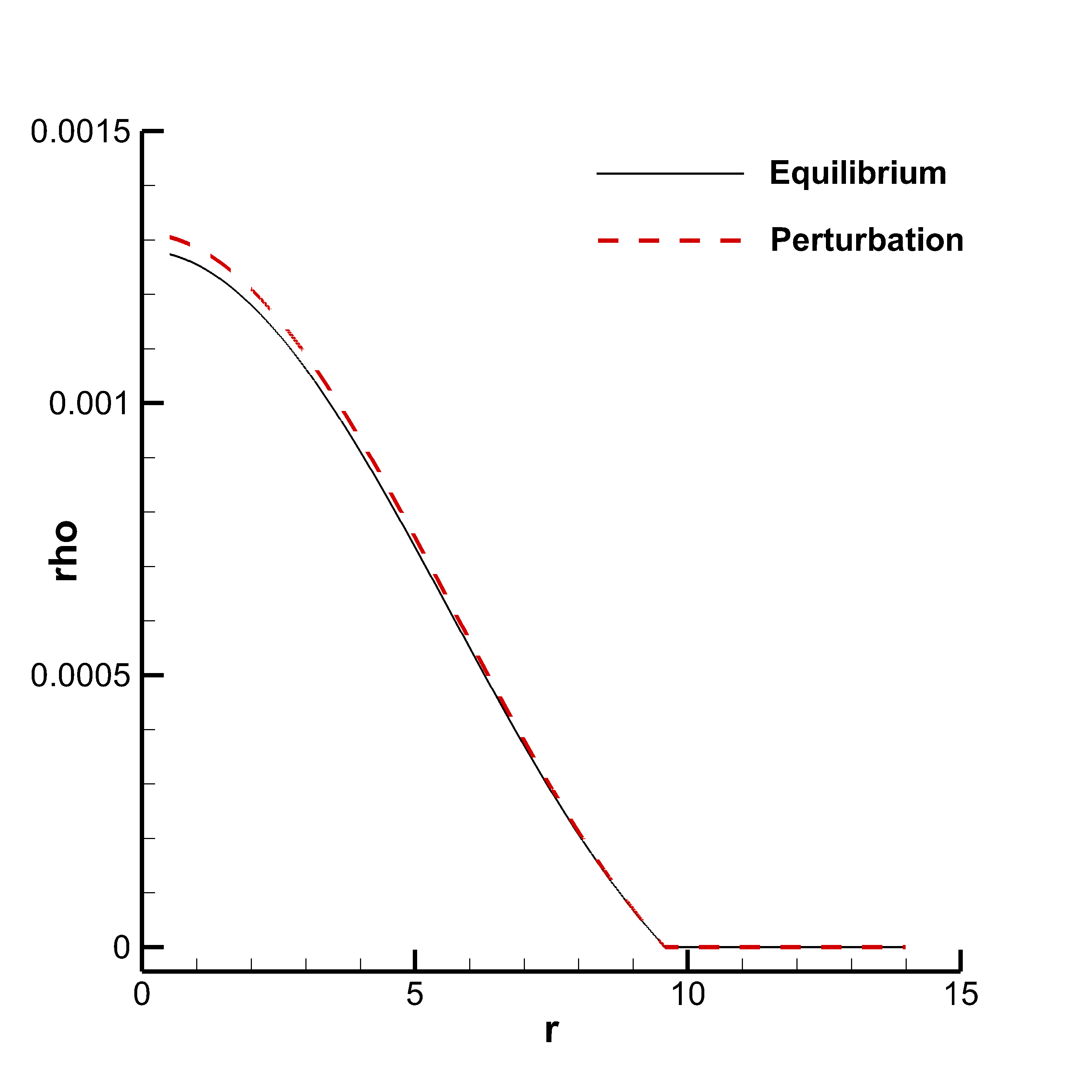}
	\includegraphics[width=0.45\linewidth]{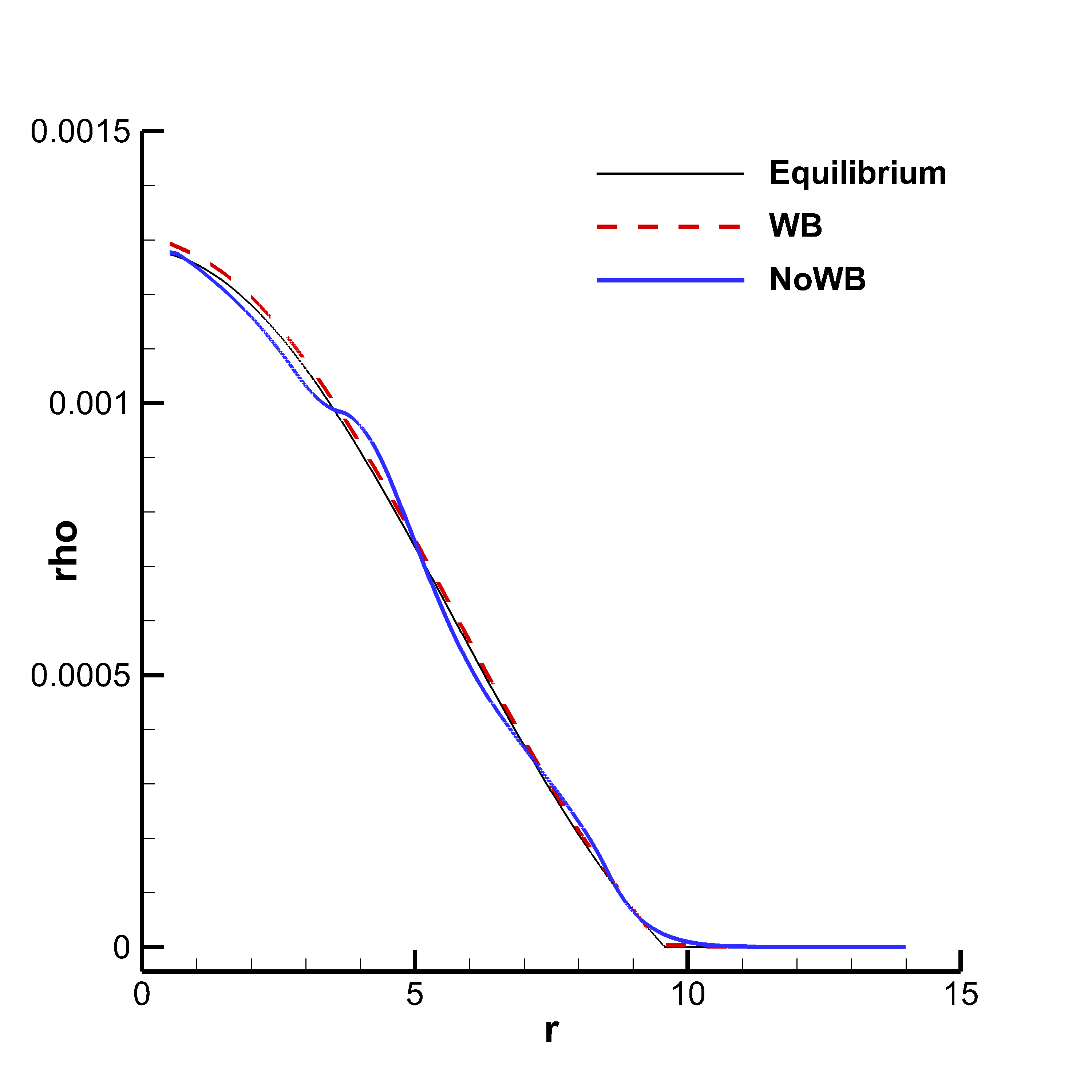}
	\vspace{-8pt}
	\caption{GRMHD TOV star simulation. 
		We show the initial pressure perturbed profile (left) and the comparison between the equilibrium profile and the numerical results obtained with the new WB and the not WB schemes at the final time $t_f = 10000$ (right). Note that only the WB algorithm is able to preserve the shape of the equilibrium aver long computational times.}
	\label{fig.grmhd}
\end{figure}
\begin{figure}
	\centering
	\includegraphics[width=0.99\linewidth]{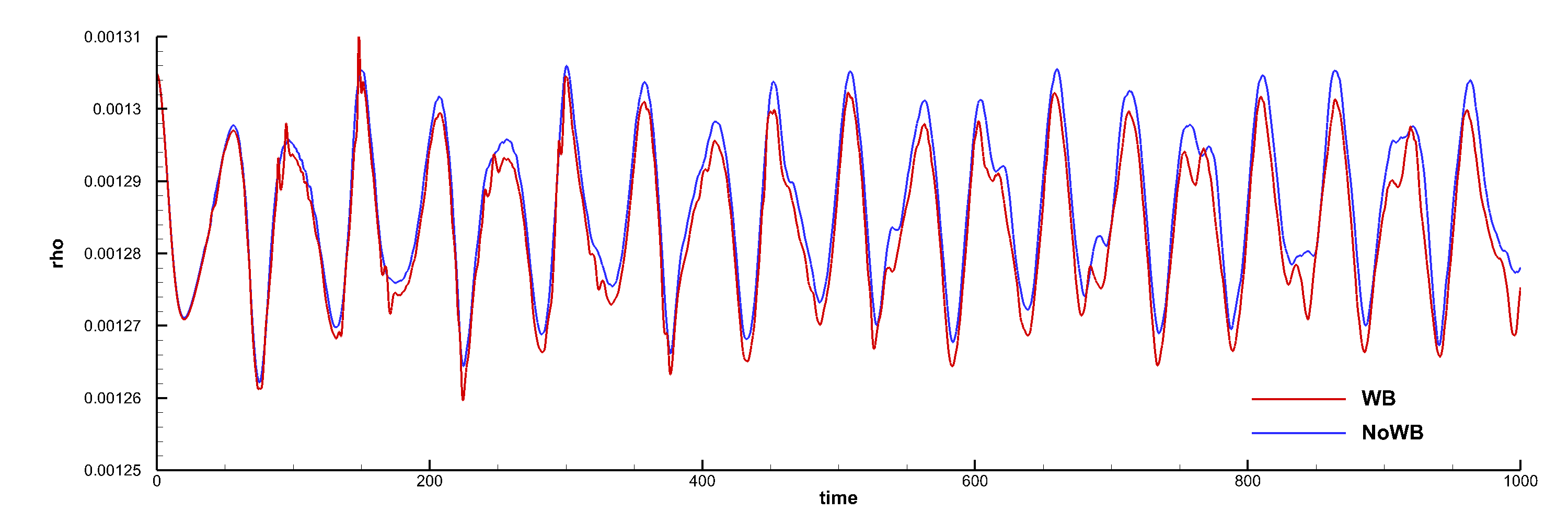}
	\vspace{-8pt}
	\caption{GRMHD TOV star simulation. We show the oscillating density of the TOV star at an inner point with $r=0.5$ over $t \in [0,1000]$ obtained with the new WB and the not WB second order schemes.}
	\label{fig.grmhdrhoc}
\end{figure}

Using the GRMHD system we can simulate the evolution of the TOV neutron star on a fixed space--time metric.
In particular, we have perturbed the initial equilibrium pressure profile as follows
\be
p = (1.0 + 0.05)p^E \quad \forall \ \x \text{ in } \Omega,
\ee 
which results also in a density perturbation, being $ \rho = (p/\kappa)^{1/\gamma}$, see the left panel of Figure~\ref{fig.grmhd}.
For what concerns the sponge layer, we have activated it only for the right boundary, 
i.e we choose $s_\mathcal{L}=0$, $s_\mathcal{R}=3$ and $\epsilon_{R}=10^{-4}$.
This allows to see the mass oscillations at the center of the star, see Figure~\ref{fig.grmhdrhoc}.

The simulation has been run with the new WB scheme and with a standard second order not WB method. 
The obtained results at the final time $t_f=10000$ are presented in Figure~\ref{fig.grmhd} 
and they show that after long simulation times the WB scheme is still able to maintain the expected profile of the solution perfectly well, while the not WB scheme is strongly affected by the accumulation of numerical errors.

%
%
%

\subsubsection{TOV neutron star simulated with FO-CCZ4 in the anti-Cowling approximation}
\label{ssec.TOVanticowling}

\begin{figure}
	\centering
	\includegraphics[width=0.49\linewidth]{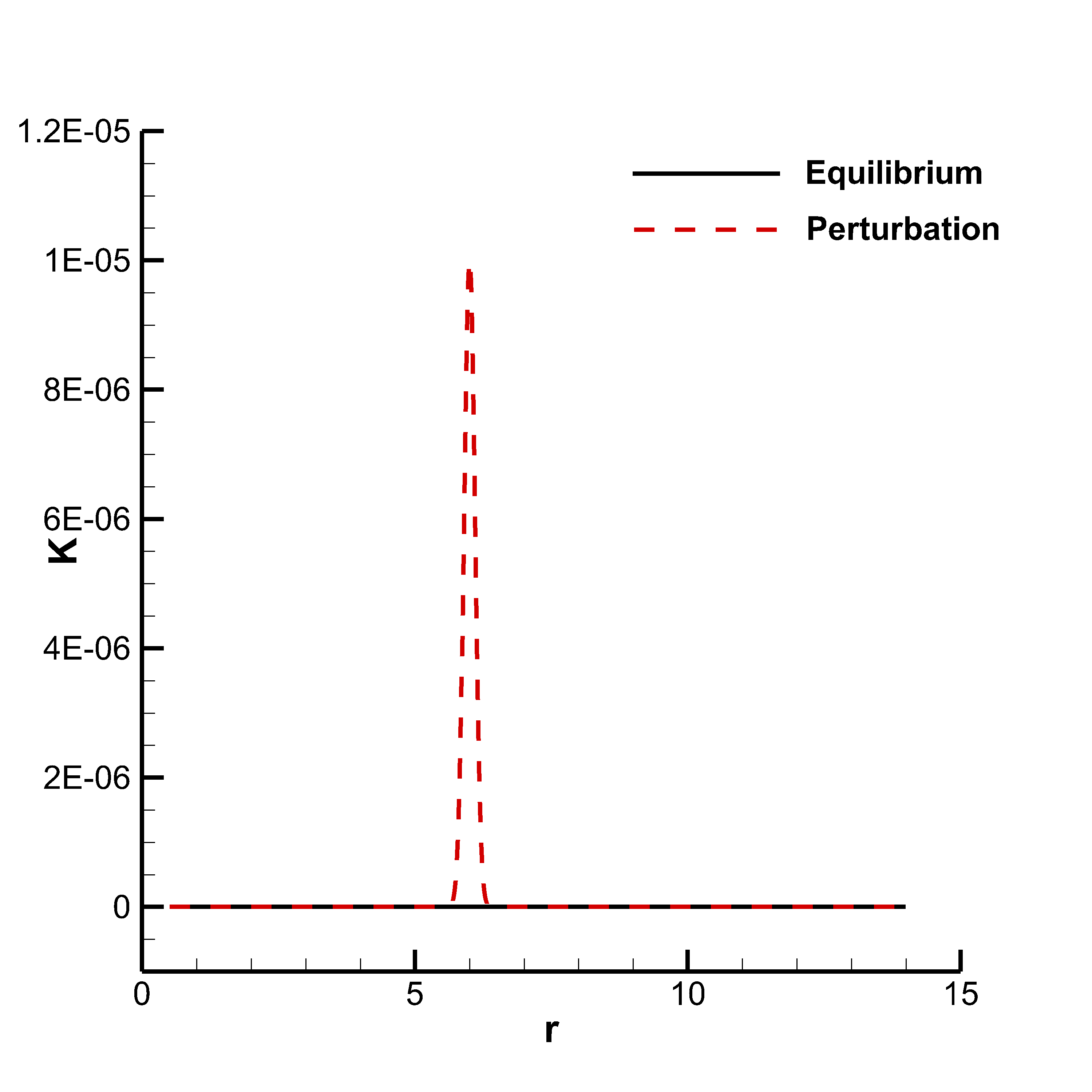}
	\includegraphics[width=0.49\linewidth]{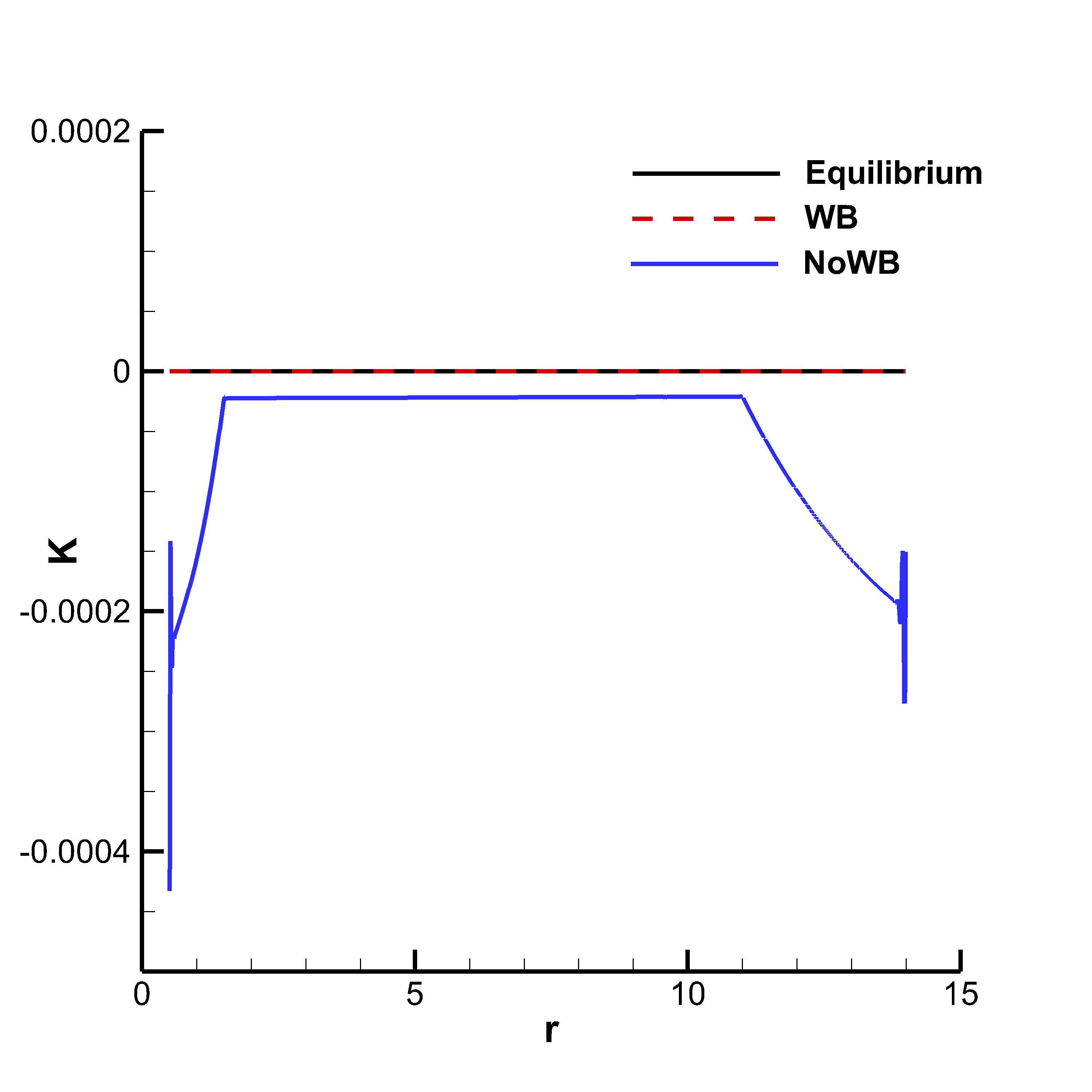}
	\vspace{-8pt}
	\caption{TOV star simulation in anti-Cowling approximation. 
		We show the initial perturbed profile of the metric variable $K$ (left) and the comparison between the equilibrium profile and the numerical results obtained with the new WB and the not WB methods at the final time $t_f = 1000$ (right). Note that the WB algorithm let the perturbation exit the domain and $K$ recovers its equilibrium profile while the not WB code corrupts the simulation. We underline that the error on the $K$ profile with the WB code are of the order of $10^{-9}$. }  
	\label{fig.anticowling_k_init}
\end{figure}
\begin{figure}
	\centering
	\includegraphics[width=0.7\linewidth]{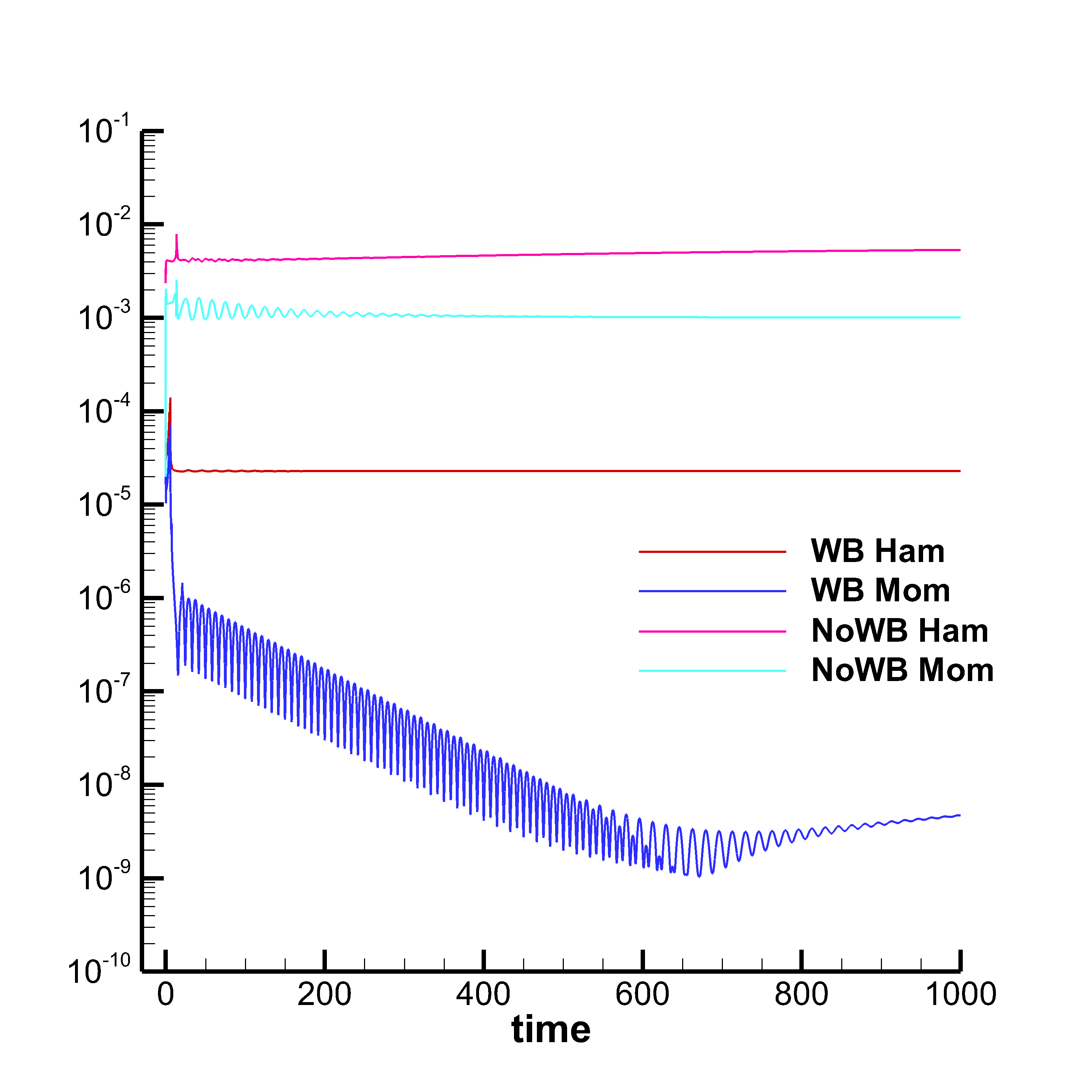}
	\vspace{-8pt}
	\caption{Hamiltonian (\texttt{Ham}) and momentum (\texttt{Mom}) constraints for the TOV star simulation in anti-Cowling approximation over the time $ t \in [0,1000]$. Note i) the initial pick due to the fact that the initial perturbed profile of $K$ does not satisfy the equations, ii) the more closer to machine precision values obtained with the WB schemes for both the constraints, iii) the decreasing/slightly increasing \blue{behaviors} of the constraints obtained with the WB/not WB codes.}
	\label{fig.anticowlingpconstraints}
\end{figure}

Next, we consider the evolution of the space--time metric in the anti-Cowling approximation,  i.e. 
we assume the matter quantities 
to be \textit{stationary} in time and externally given 
by the Tolman-Oppenheimer-Volkoff (TOV) solution. 

As gauge conditions we employ a frozen shift condition $\partial_t \beta^i = 0$ by setting $s=0$ in the FO-CCZ4 system, 
together with the harmonic slicing, which corresponds to the choice $g(\alpha)=1$. The cleaning speed for the cleaning of the nonlinear ADM constraints is set to $e=1.5$.
The remaining constants in the FO-CCZ4 system are chosen as $\gamma=1.4$, $\kappa_1=\kappa_2=\kappa_3=0$, $c=0$ and $\eta=0$.  

We study here the effects of an initial perturbation over the variable $K$ so that 
\be
\label{eq.Pert_K}
K(r,0) = K^E(r) + 10^{-5} \exp \left ( -0.5 \frac{(r-6.0)^2}{0.1^2} \right ),
\ee 
see also the left panel of Figure~\ref{fig.anticowling_k_init}. 
The bump should split into two waves which are expected to propagate until exiting the domain. 
This happens when using the WB scheme where indeed the equilibrium profile of $K$ is soon recovered  (with an error of $10^{-9}$ at the final time $t_f =1000$)
while with a not WB scheme the presence of numerical errors prevents the restoring of the equilibrium, see the right panel of Figure ~\ref{fig.anticowling_k_init}.
For what concerns the sponge layer, 
we choose $s_\mathcal{L}=1.0$, $s_\mathcal{R}=3$, $\epsilon_{L}=10^{-3}$ and $\epsilon_{R}=10^{-4}$ to avoid the spurious reflection of waves at the boundaries.

Finally, we monitor the Hamiltonian and momentum constraints~\eqref{eq.constraints} over the time, refer to Figure~\ref{fig.anticowlingpconstraints}.
Since the initial perturbed profile of $K$~\eqref{eq.Pert_K} does not satisfy the equations we can notice a higher peak at the beginning of the simulation.
Then, once the perturbation has exited the domain, the WB scheme is able to recover constraint values closer to zero w.r.t to the not WB scheme.

\subsubsection{TOV neutron star simulated with the fully coupled FO-CCZ4 $+$ GRHD model}
\label{ssec.TOVfullycoupled}

Finally, we consider the fully coupled model that can be obtained by considering together the FO-CCZ4 model 
and the GRMHD model; the initial metric values and the parameters for the ADM constraints are set as in the previous test case.
In what follows we study two different perturbations of the stationary TOV neutron star profile.
We would like to remark that, up to our knowledge, these are the \textit{first} numerical simulations ever presented with a well balanced finite volume scheme for the fully coupled Einstein-Euler system. 
Long time stability is achieved thanks to the use of the novel WB techniques, which are able to stabilize the simulation and avoid the 
accumulation of spurious numerical errors.

\paragraph{\bf Perturbation of the metric variable $\mathbf{K}$}
\begin{figure}
	\centering 
	\includegraphics[trim= 0 15 0 35,clip,width=0.49\linewidth]{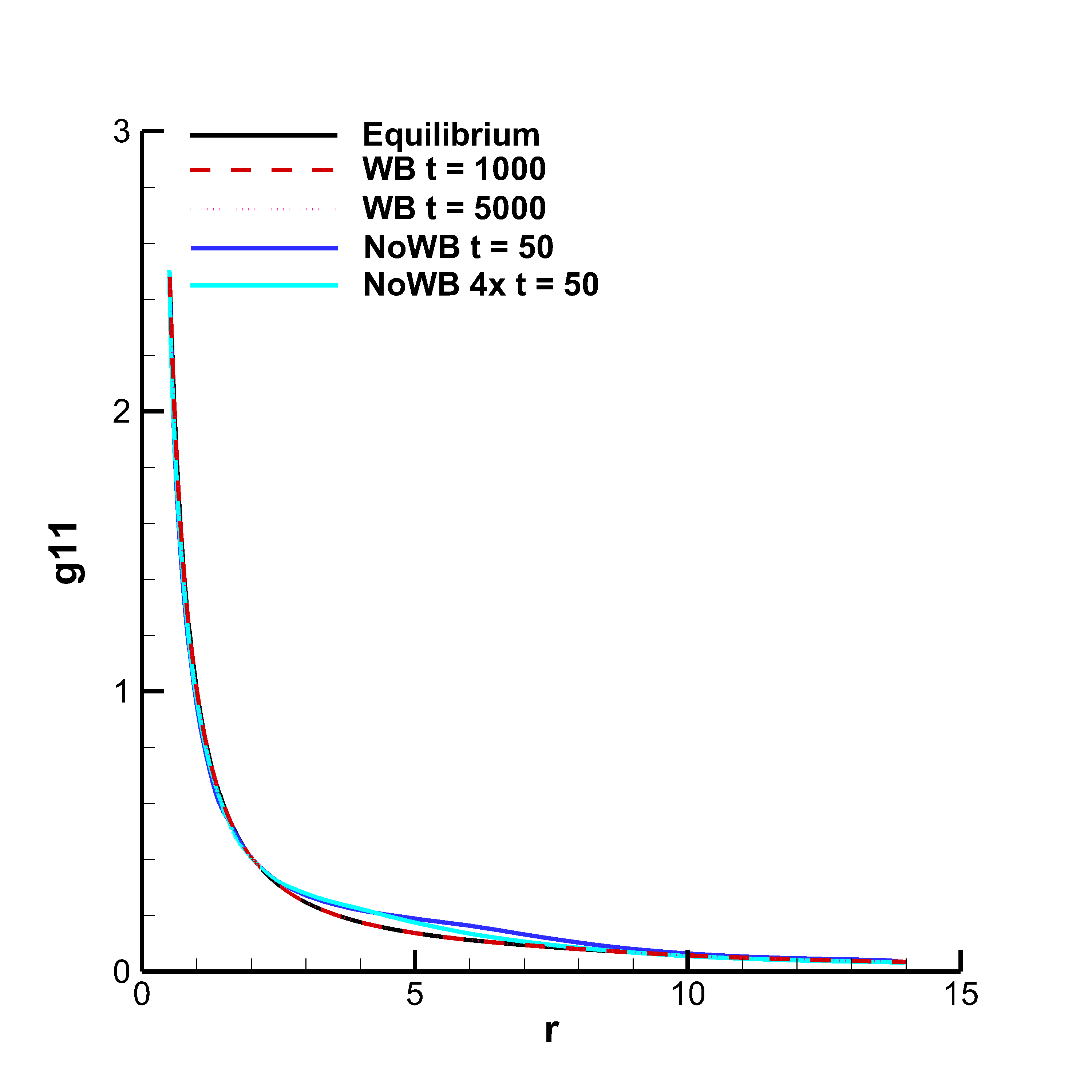}
	\includegraphics[trim= 0 15 0 35,clip,width=0.49\linewidth]{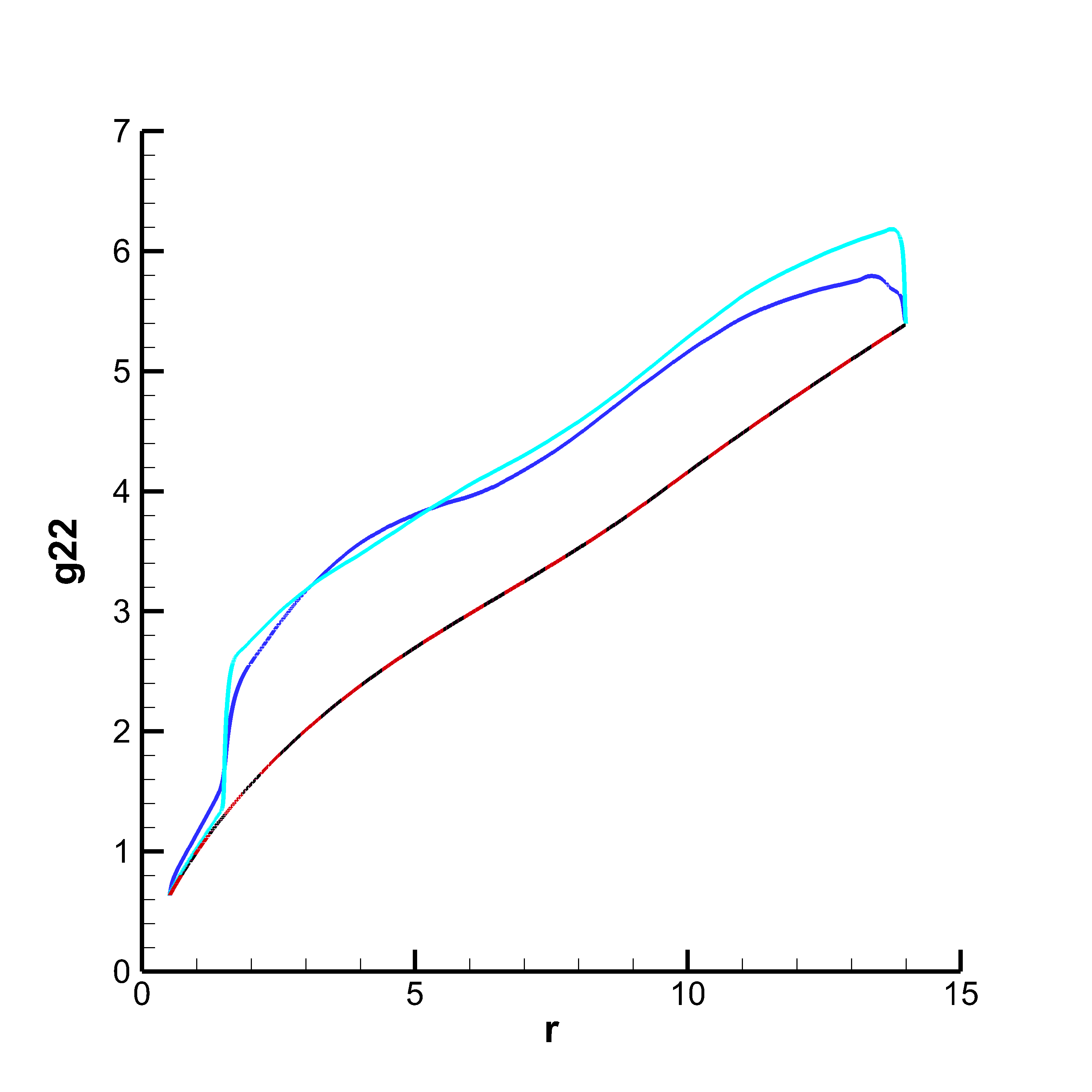}
	\includegraphics[trim= 0 15 0 35,clip,width=0.49\linewidth]{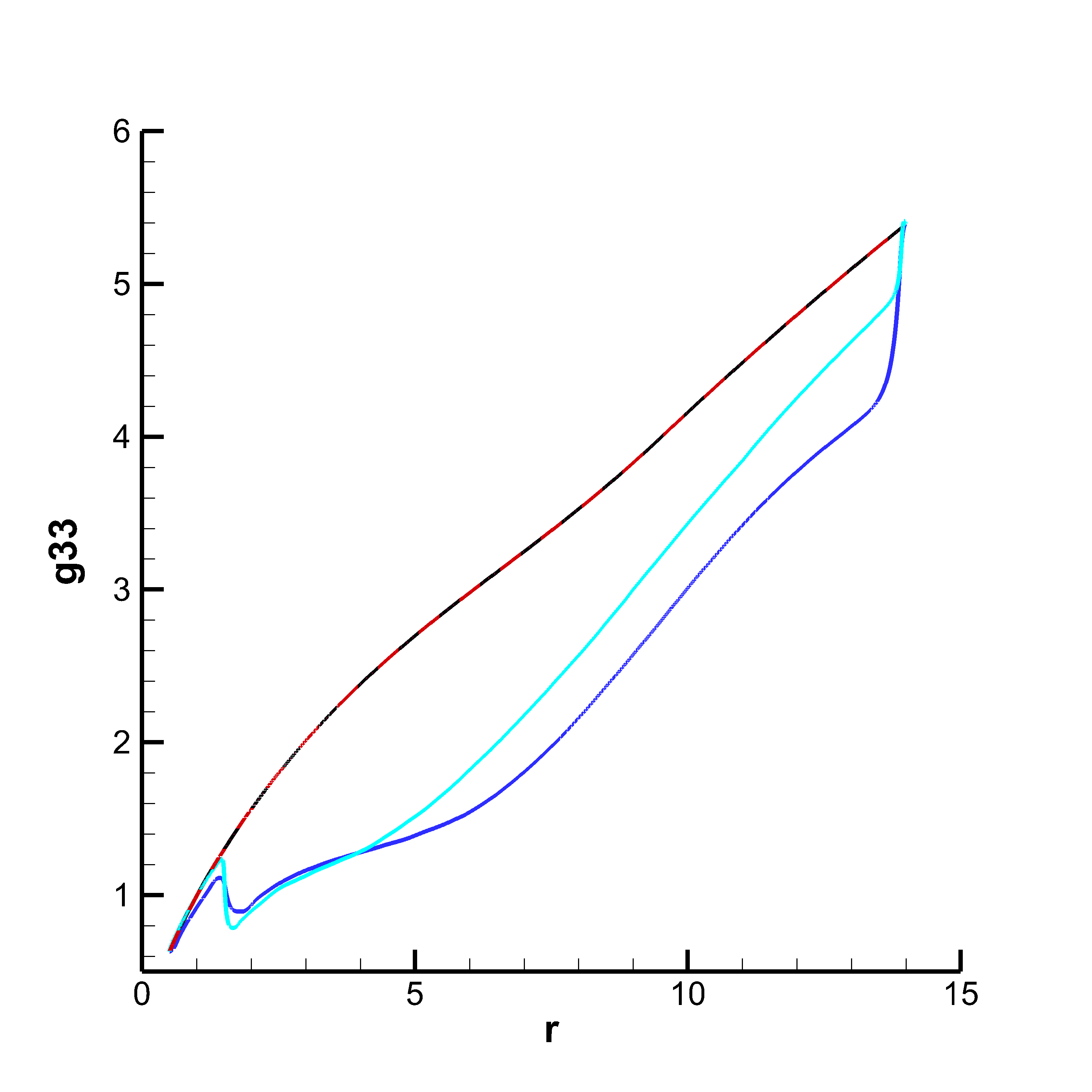}
	\includegraphics[trim= 0 15 0 35,clip,width=0.49\linewidth]{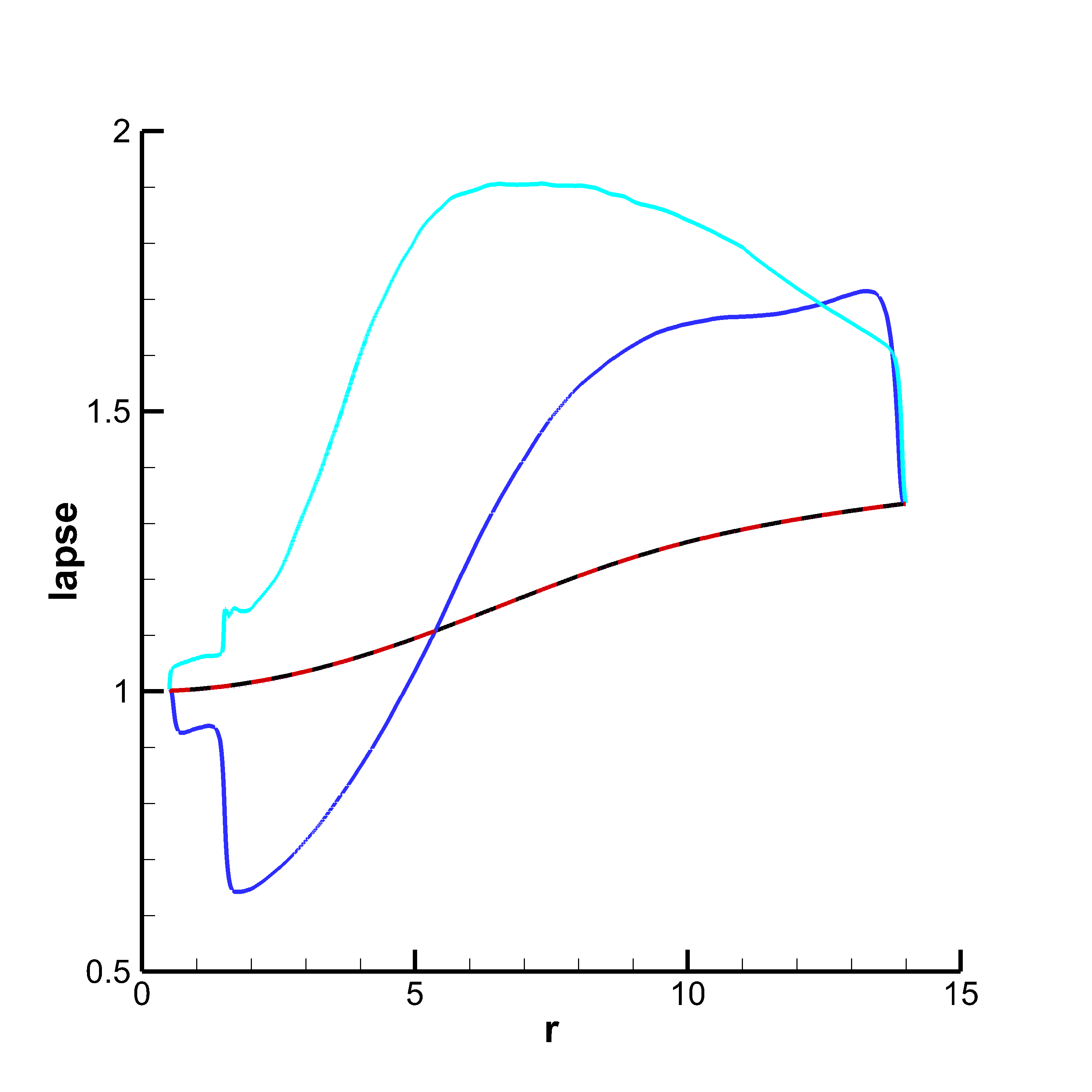}
	\includegraphics[trim= 0 15 0 35,clip,width=0.49\linewidth]{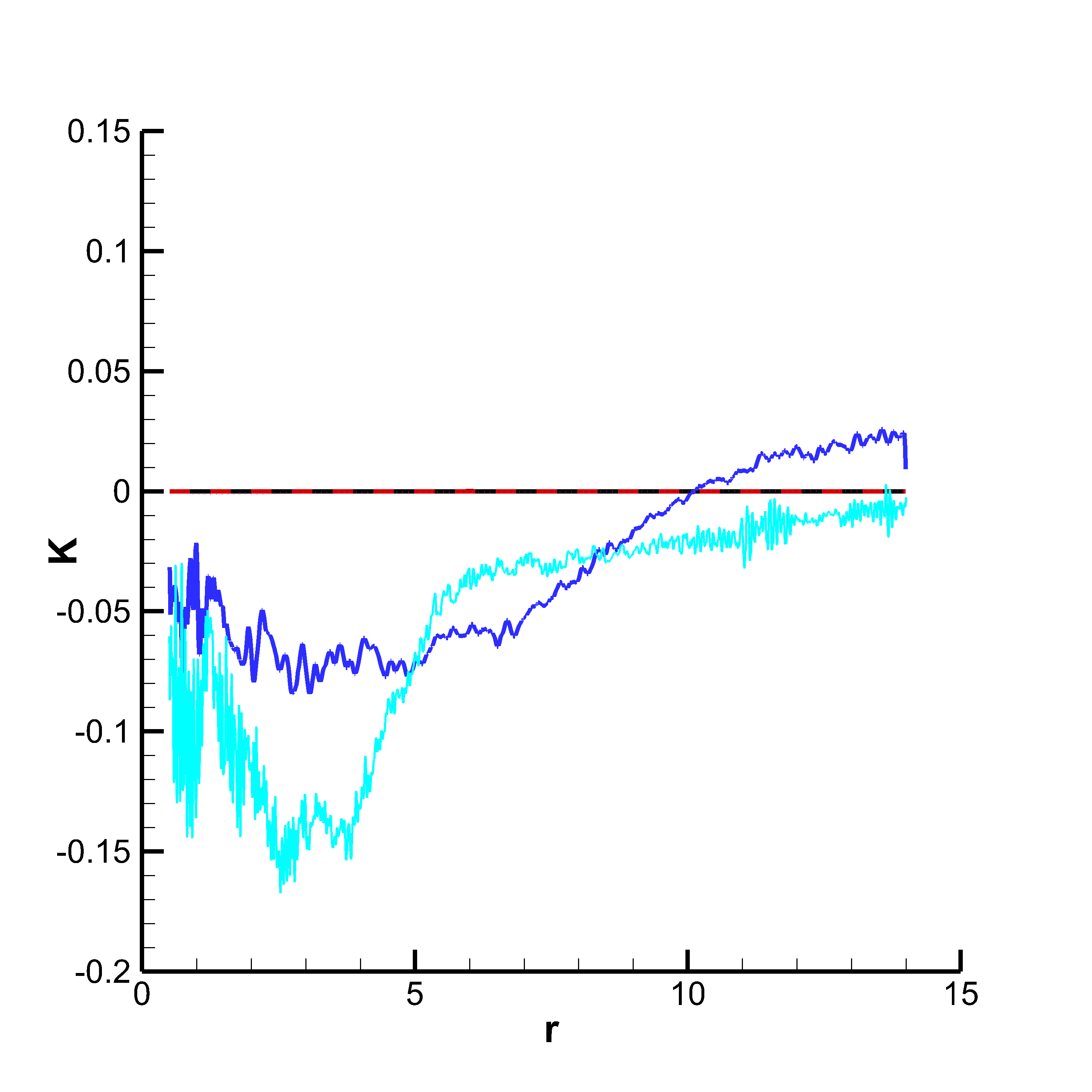}
	\includegraphics[trim= 0 15 0 35,clip,width=0.49\linewidth]{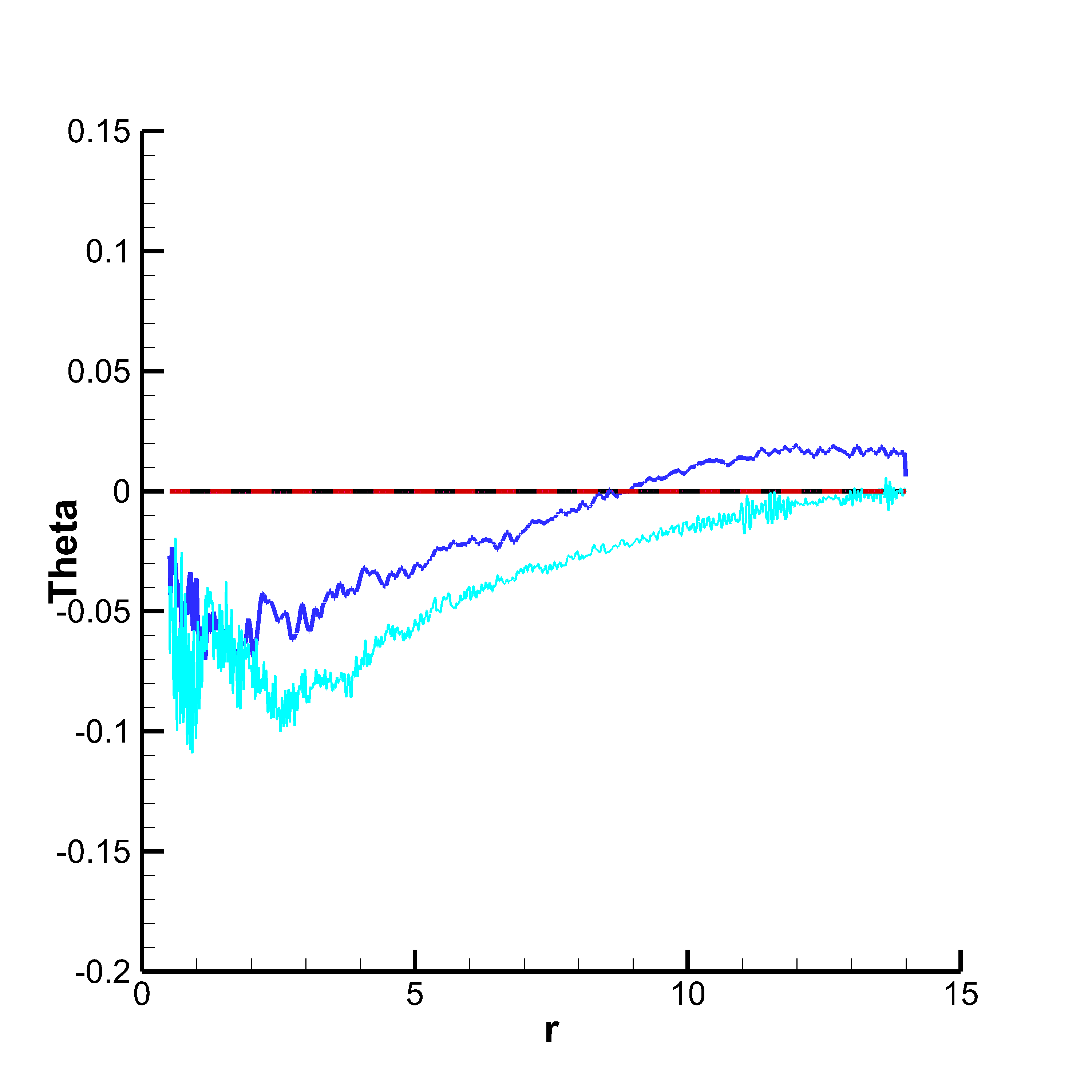}
	\vspace{-8pt}
	\caption{TOV star simulation with the CCZ4+GRHD fully coupled model corresponding to an initial perturbation over the metric variable $K$. We show: i) the expected equilibrium profile to be recovered once the perturbation has exited the domain (black line), ii) the numerical results obtained with the new WB scheme at very long times ($t=1000$ in red and $t=5000$ in pink) and iii) the completely corrupted results obtained with the not WB scheme already after a very short computational time $t=50$ (blue) even on a four time finer mesh (cyan). Plotted variables: the diagonal components of the metric $\gamma_{11}$, $\gamma_{22}$, $\gamma_{33}$, the lapse function $\lambda$, the extrinsic curvature $K$ and the cleaning variable~$\theta$.  }
	\label{fig.fullycoupled_k_Metric}
\end{figure}

\begin{figure}
	\centering
	\includegraphics[width=0.49\linewidth]{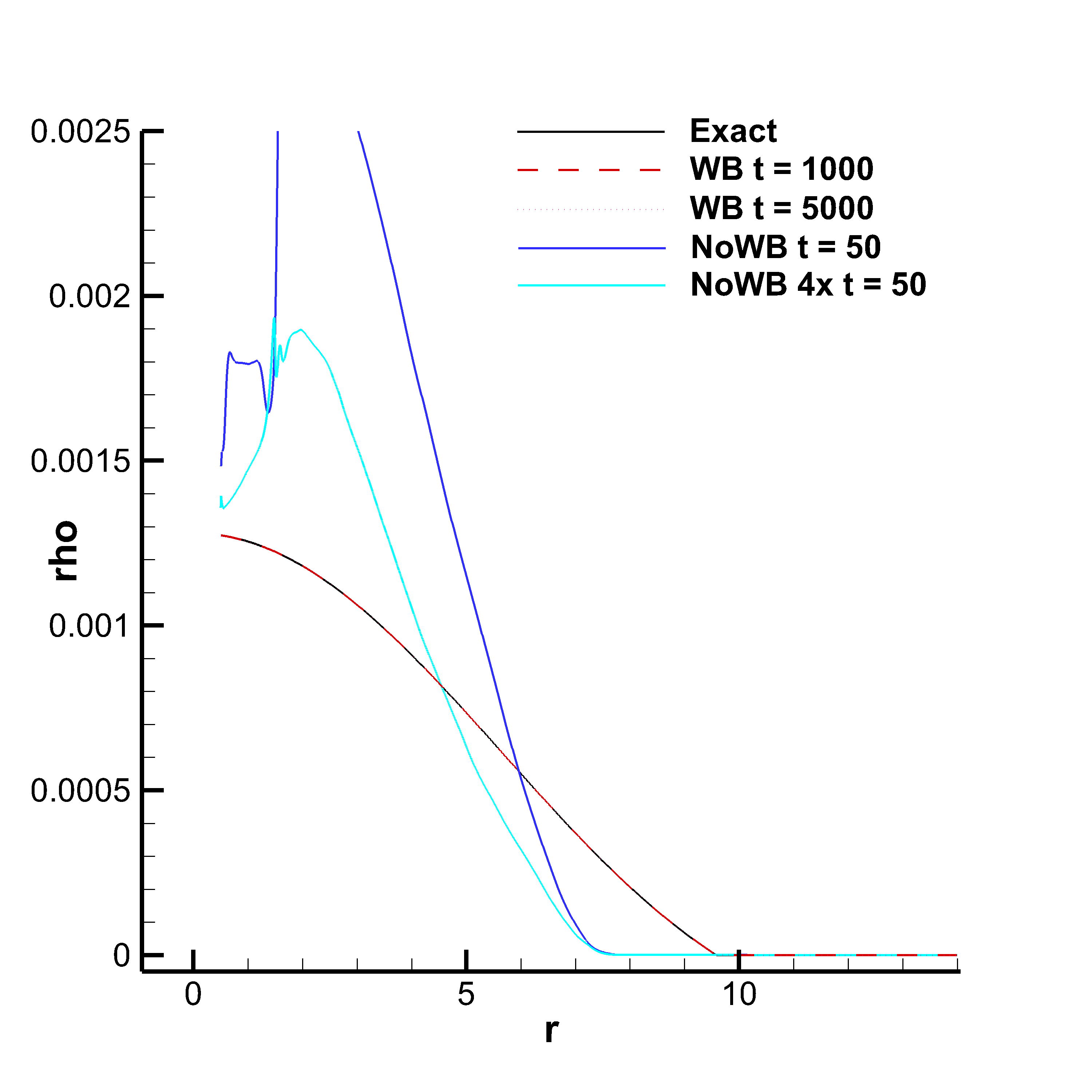}
	\includegraphics[width=0.49\linewidth]{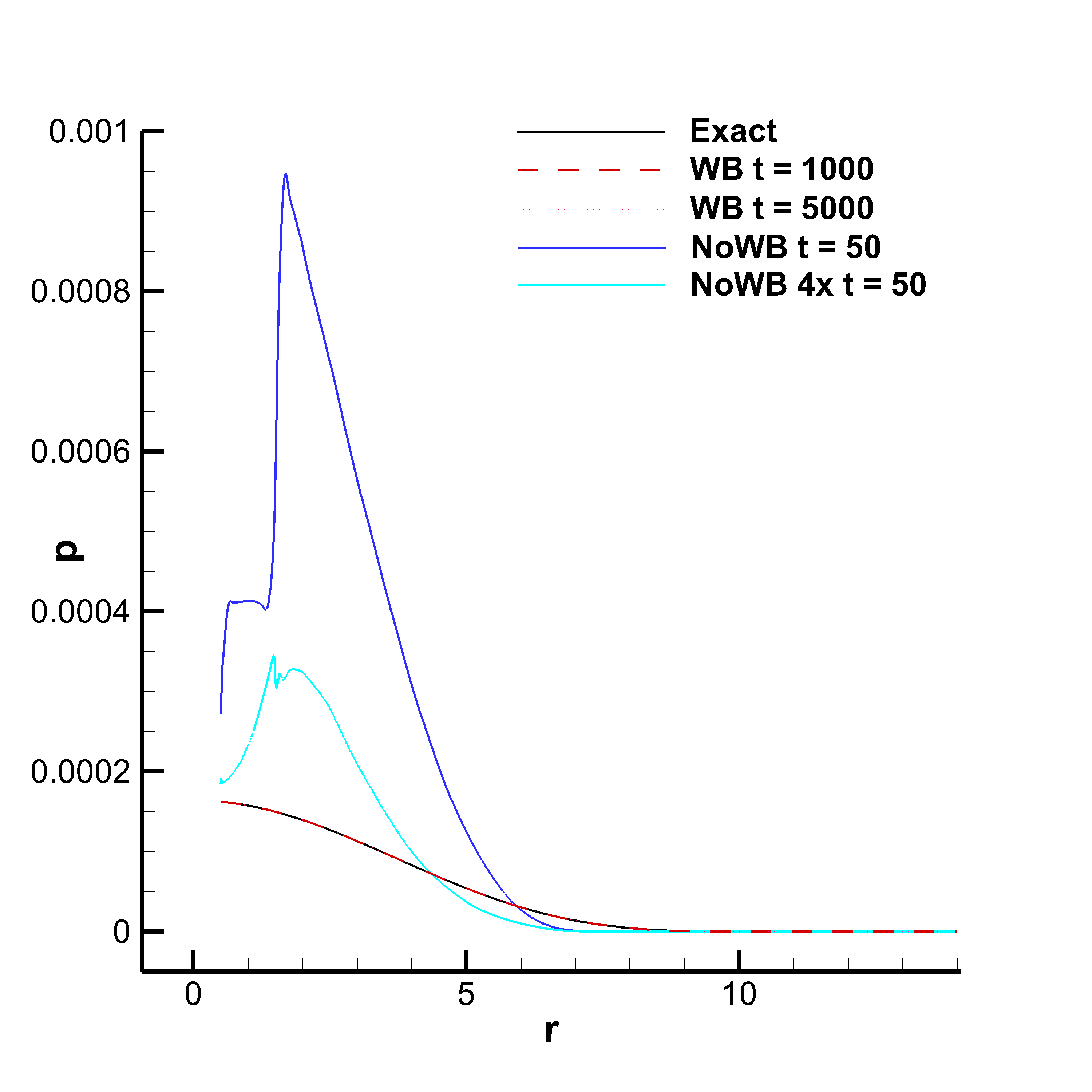}
	\vspace{-10pt}
	\caption{TOV star simulation with the CCZ4+GRHD fully coupled model corresponding to an initial perturbation over the metric variable $K$. With the same color map of the previous Figure we show here the effects of a perturbation over the metric variable $K$ on matter variables as density $\rho$ (left) and pressure $p$ (right).}
	\label{fig.fullycoupled_k_Matter}
\end{figure}

\begin{figure}
	\centering
	\includegraphics[width=0.7\linewidth]{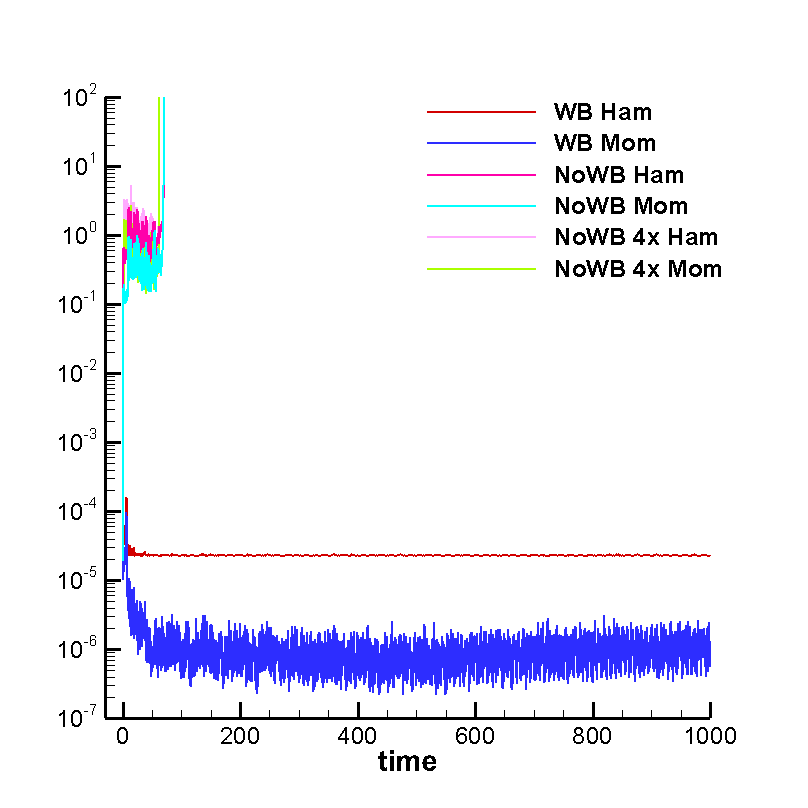}
	\vspace{-10pt}
	\caption{Hamiltonian (\texttt{Ham}) and momentum (\texttt{Mom}) constraints for the TOV star simulation with the fully coupled model over the time $t \in [0,1000]$. Note again the initial pick, the more closer to machine precision values obtained with the WB schemes already at the beginning of the simulation and in particular the fact that with the not WB code the constraints, as the entire simulation, completely blows up after a brief time while the WB code always provided trustable numerical results as evident from the perfect constraints preservation.
	}
	\label{fig.fullycoupled_k_Constraints}
\end{figure}

First, we perturb the initial value of the metric variable $K$ as in~\eqref{eq.Pert_K} and for what concerns the sponge layer, 
we choose $s_\mathcal{L}=1.0$, $s_\mathcal{R}=3$, $\epsilon_{L}=10^{-3}$ and $\epsilon_{R}=10^{-4}$.

We report in Figure~\ref{fig.fullycoupled_k_Metric} the profile of some metric variables and in Figure~\ref{fig.fullycoupled_k_Matter} the profile of the density and pressure in the matter. 
These results have been obtained on one hand by using the new WB scheme presented in this paper, which allows to recover and maintain the equilibrium profile even for very long simulation times of $t=1000$ or $t=5000$, and on the other hand with a standard second order scheme
that, even on a four times finer mesh completely destroys the solution profile in a time of only $t=50$. 

The difference in resolution and stability between the WB and not WB scheme is made evident also by the monitoring of the Hamiltonian and momentum constraints, see Figure~\ref{fig.fullycoupled_k_Constraints}: right from the beginning, the WB scheme allows a much more precise representation of the discrete solution and consequently also the errors in the involution constraints are much lower; moreover, these smaller values of $\mathcal{H}$ and $\mathcal{M}$ are almost constantly maintained in time by the WB scheme, while their values explode soon with the not WB scheme.

\paragraph{\bf Perturbation of the matter variable $\mathbf{p}$}

\begin{figure}
	\centering
	\includegraphics[width=0.49\linewidth]{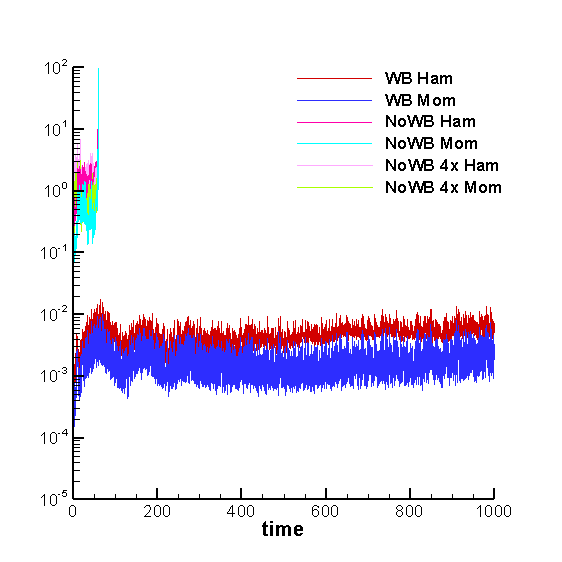}
	\includegraphics[width=0.49\linewidth]{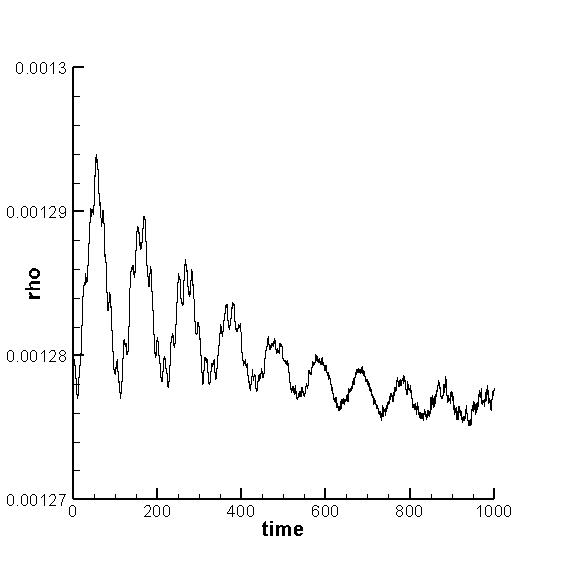}
	\vspace{-10pt}
	\caption{TOV star simulation with the CCZ4+GRHD fully coupled model corresponding to an initial perturbation over the pressure profile $p$. Left: we show the values of the Hamiltonian and momentum constraints which are close to machine precision and constantly maintained for long computational times with the WB code, while they rapidly blow up with the not WB code. Right: we plot the numerical results obtained with the new WB scheme for the density of the star at an inner point with $r=0.5$ which oscillates due \blue{to} the initial pressure perturbations.}
	\label{fig.fullycoupledConstraints_rhoc_Ppert}
\end{figure}

Next, we study the effect of a small pressure perturbation as the one given in~\eqref{eq.Pert_P} where, for what concerns the sponge layer,
we activate it only for the right boundary, and we choose $s_\mathcal{L}=0$, $s_\mathcal{R}=3$ and $\epsilon_{R}=10^{-4}$.
Once again the constraints are more precise and better preserved with the WB scheme, see Figure~\ref{fig.fullycoupledConstraints_rhoc_Ppert}.

For what concerns the mass oscillation at an internal point with $r=0.5$ in Figure~\ref{fig.fullycoupledConstraints_rhoc_Ppert} we report only 
the results obtained with the new WB scheme, since the not WB method, even on finer meshes, explodes after a very short time.  
Comparing the amplitude of this oscillation with those obtained on a fixed space--time in Figure~\ref{fig.grmhdrhoc} one can appreciate the different 
dynamic behavior obtained with a fully coupled model, where the interaction between metric and mass are taken into account, compared to the simpler GRMHD model in Cowling approximation with fixed space-time.   

\blue{ 
	\subsection{Numerical convergence studies for our well balanced algorithm}
	\label{ssec.convergence}
	
	For the sake of completeness, in this last section we check the convergence rate of our new well balanced scheme by computing the experimental errors committed 
	when solving the Michel accretion disk test case, which is an equilibrium solution of the GRMHD system known analytically (see Section~\ref{ssec.AccretionDisk}),
	while preserving completely different equilibrium profiles, namely a i) fluid at rest immersed in a flat space-time
	and ii) the TOV neutron star of the previous Section~\ref{ssec.TOVStar}.  
	The choice of using an initial condition that completely differs from the equilibrium profiles to be preserved 
	allows to show the second order of convergence of our well balanced method; the scheme would be instead \textit{exact} when choosing to preserve 
	an equilibrium that is equal to the initial condition of the test problem. 
	The obtained numerical results are shown in Table~\ref{tab.orderMichelOnFlatST}; 
	we would like to emphasize that the order of convergence is correctly retained even for long computational times.

	\begin{table}\blue{ 
			\caption{$L_2$ norm of the experimental errors and order of accuracy at times $t=1,10,100$ and $1000$ for the density variable $\rho$ of the Michel accretion disk test case simulated with our well balanced scheme set to preserve i) a fluid at rest over a flat space-time (top) and ii)
				a TOV Neutron star (bottom).} 
			\label{tab.orderMichelOnFlatST}
			\begin{center} 	
				\begin{tabular}{|c|cc|cc|cc|cc|} 
					\hline 
					Flat ST  &     \mcol{$t=1$}      &     \mcol{$t=10$}    &    \mcol{$t=100$}    &    \mcol{$t=1000$}    \\[1pt]
					$\Delta x$   &   $L_2$   &  \OLdue   &  $L_2$  &  \OLdue    &  $L_2$   &  \OLdue   &  $L_2$   &  \OLdue    \\[1pt]
					\hline 			
					9.0E-03      &  3.1E-06  &   -       & 4.9E-05 &   -        &  5.9E-04  &  -       & 5.0E-03  &    -       \\[1pt]
					6.0E-03      &	1.4E-06  &   1.99    & 2.2E-05 &   1.93     &  2.7E-04  &  1.88    & 2.4E-03  &    1.79    \\[1pt]
					4.5E-03      &  8.0E-07  &   1.95    & 1.3E-05 &   1.95     &  1.6E-04  &  1.91    & 1.4E-03  &    1.82    \\[1pt]
					3.6E-03      &  5.1E-07	 &	 1.98    & 8.3E-06 &   1.96     &  1.0E-04  &  1.92    & 9.5E-04  &    1.84    \\[1pt]
					3.0E-03      &  3.6E-07  &   1.97    & 5.8E-06 &   1.96     &  7.3E-05  &  1.93    & 6.8E-04  &    1.86   \\[1pt]
					\hline  
					TOV star &     \mcol{$t=1$}      &     \mcol{$t=10$}    &    \mcol{$t=100$}    &    \mcol{$t=1000$}    \\[1pt]
					$\Delta x$   &   $L_2$   &  \OLdue   &  $L_2$  &  \OLdue    &  $L_2$   &  \OLdue   &  $L_2$   &  \OLdue    \\[1pt]
					\hline 
					9.0E-03      &  2.7E-06  &   -       & 5.6E-05 &   -        &  7.2E-04  &  -       & 6.3E-03  &    -      \\[1pt]
					6.0E-03      &	1.2E-06  &   1.99    & 2.6E-05 &   1.92     &  3.3E-04  &  1.89    & 3.0E-03  &    1.81   \\[1pt]
					4.5E-03      &  6.9E-07  &   1.94    & 1.5E-05 &   1.94     &  1.9E-04  &  1.91    & 1.8E-03  &    1.84   \\[1pt]
					3.6E-03      &  4.4E-07	 &	 1.98    & 9.6E-06 &   1.95     &  1.3E-04  &  1.92    & 1.2E-03  &    1.86   \\[1pt]
					3.0E-03      &  3.1E-07  &   1.97    & 6.7E-06 &   1.96     &  8.8E-05  &  1.93    & 8.4E-04  &    1.87   \\[1pt]
					\hline
				\end{tabular}		
		\end{center}  }
	\end{table}
}

\section{Conclusions}
\label{sec.Conclusions}

With the numerical results presented here we have clearly shown the advantages that can be obtained in numerical general relativity 
through the use of a simple but very efficient well balancing technique.  
Indeed, simulations that blow up with a standard second order scheme 
can  now be carried out in a stable manner even for very long simulation times and with accurate results, 
as demonstrated for example by the monitoring of the ADM constraints and the simulations carried out with the fully coupled FO-CCZ4 + GRMHD model.  
Moreover, we would like to underline that this preliminary work in 1D at second order of accuracy has been essential to
understand model related difficulties how the setting of the cleaning and damping parameters in the FO-CCZ4 model and the boundary conditions.

Future work will concern the insertion of the presented WB techniques in already existing higher order discontinuous Galerkin schemes for modeling the GRMHD and the FO-CCZ4 systems in two or three space dimensions, as those presented in~\cite{Dumbser2018conformal,Gaburro2021PNPMLimiter}. In this context we will follow also the numerical approach presented in~\cite{xing2006high,xing2014exactly}. 
Moreover, we would like to investigate the joint effect of WB and the novel GLM curl cleaning techniques introduced in~\cite{Dumbser2020GLM,chiocchetti2021high,HyperbolicDispersion}. \blue{Indeed, several systems with curl involutions have recently been investigated and attention has been devoted to the crucial role of curl involutions themselves \cite{dumbser2020involutions} on the stability of numerical computations. 
With the increased level of robustness and accuracy that can be obtained at the aid of the new well balanced schemes introduced in this paper, in combination with high order DG schemes and novel curl cleaning techniques,} it should therefore become possible to study also the generation and propagation of \textit{gravitational waves} in the future. 

Furthermore, motivated by the results obtained in~\cite{Gaburro2018MNRAS} for the study of Keplerian disks 
modeled with the Euler equations with gravity and simulated with a well balanced \textit{Lagrangian} scheme,  
we would like to apply a similar approach also for the study of general relativistic phenomena; 
in particular, we plan to incorporate our new well balanced techniques for GRMHD and FO-CCZ4 
in modern direct Arbitrary-Lagrangian Eulerian algorithms with topology changes, as those forwarded in~\cite{Springel,Gaburro2017CAFNonConf,Gaburro2020Arepo,Gaburro2020,cirrottola2021adaptive}. 

\section*{Acknowledgments}

E.~G gratefully acknowledges the support received from the European Union’s Horizon 2020 Research and Innovation Programme under the Marie Skłodowska-Curie Individual Fellowship \textit{SuPerMan}, grant agreement No. 101025563.
E.~G. has been also supported by a national mobility grant for young researchers in Italy funded by GNCS-INdAM, 
and she received funding from the University of Trento via the Strategic Initiative \textit{Starting Grant Giovani Ricercatori 2019}.

Moreover, the research presented in this paper has been partially funded by the European Union's Horizon 2020 Research and Innovation Programme under the project \textit{ExaHyPE}, grant No. 671698 (call FETHPC-1-2014). 

M.~D. also acknowledges the financial support received from 
the Italian Ministry of Education, University and Research (MIUR) in the frame of the Departments of Excellence  Initiative 2018--2022 attributed to DICAM of the University of Trento (grant L. 232/2016) and in the frame of the 
PRIN 2017 project \textit{Innovative numerical methods for evolutionary partial differential equations and applications}. 
Furthermore, M.~D. has also received funding from the University of Trento via the Strategic Initiative \textit{Modeling and Simulation}. 

The  research  of M.J. Castro  was  partially  supported  by  the  Spanish  Government (SG),  
the  European  Regional  Development Fund (ERDF), the Regional Government of Andalusia (RGA), 
and the University of M\'alaga(UMA) through the projects  of  references  RTI2018-096064-B-C21  (SG-ERDF),  UMA18-Federja-161  (RGA-ERDF-UMA), and P18-RT-3163 (RGA-ERDF).

E.G. is member of the CARDAMOM team at Inria BSO (France), M.D. is member of the INdAM-GNCS group (Italy), and M. J. Castro is member of the EDANYA group (Spain). 

\bibliographystyle{mysiamplain}  
\bibliography{referencesWBGR2}
\end{document}